\documentclass[english,american]{article}

\usepackage[T1]{fontenc}
\usepackage[latin9]{inputenc}
\usepackage[a4paper]{geometry}
\usepackage{fancyhdr}
\usepackage{babel}
\usepackage{array}
\usepackage{prettyref}
\usepackage{multirow}
\usepackage{amsmath}
\usepackage{amssymb}
\usepackage{graphicx}
\usepackage[authoryear]{natbib}
\usepackage{xargs}[2008/03/08]
\usepackage[unicode=true,
 bookmarks=true,bookmarksnumbered=false,bookmarksopen=false,
 breaklinks=true,pdfborder={0 0 0},pdfborderstyle={},backref=false,colorlinks=false]
 {hyperref}
\usepackage{breakurl}

\makeatletter

\providecommand{\tabularnewline}{\\}

\newrefformat{alg}{Alg.~\ref{#1}}
\newrefformat{fig}{Fig.~\ref{#1}}
\newrefformat{tab}{Table~\ref{#1}}
\newrefformat{sec}{Sec.~\ref{#1}}
\newrefformat{sub}{Sec.~\ref{#1}}
\newrefformat{subsec}{Sec.~\ref{#1}}
\newrefformat{par}{Sec.~\ref{#1}}
\newrefformat{apx}{Appendix~\ref{#1}}

\@ifundefined{definecolor}
{\usepackage{color}}{}

\definecolor{parametergray}{gray}{0.8}

\usepackage[OT1]{fontenc}
\usepackage{colortbl}
\usepackage{calc}

\renewenvironment{abstract}
{\noindent{\normalfont\large\textbf{Abstract}}%
\par\vspace{0.5\baselineskip}}
{\par}

\renewcommand{\@seccntformat}[1]{%
\csname the#1\endcsname\hspace{0.5em}}

\renewcommand{\section}{\@startsection
{section}%
{1}%
{0mm}%
{-\baselineskip}%
{0.5\baselineskip}%
{\normalfont\large\bfseries}}

\renewcommand{\subsection}[1]{\ssubsection{#1.}}
\newcommand{\ssubsection}{\@startsection
{subsection}%
{2}%
{1em}%
{-\baselineskip}%
{-\fontdimen2\font plus -\fontdimen3\font minus -\fontdimen4\font}%
{\normalfont\bfseries}}

\renewcommand{\subsubsection}[1]{\sssubsection{#1.}}
\newcommand{\sssubsection}{\@startsection
{subsubsection}%
{3}%
{1em}%
{-\baselineskip}%
{-\fontdimen2\font plus -\fontdimen3\font minus -\fontdimen4\font}%
{\normalfont\itshape}}

\renewcommand{\paragraph}[1]{\ssssubsection{#1.}}
\newcommand{\ssssubsection}{\@startsection
{paragraph}%
{4}%
{1em}%
{-\baselineskip}%
{-\fontdimen2\font plus -\fontdimen3\font minus -\fontdimen4\font}%
{\normalfont\itshape}}

\renewcommand{\@makecaption}[2]{%
{\parbox[t]{\linewidth}{%
\normalsize\renewcommand{\baselinestretch}{1.8}\normalsize
\vspace{2mm}
\textbf{#1:} #2
}}}

\usepackage{babel}

\usepackage{subfig}
\usepackage{subfig}
\usepackage{subfig}

\usepackage{dsfont}
\date{}

\makeatother

\begin{document}
\global\long\def\ms{\:\mathrm{ms}}
\global\long\def\mV{\:\mathrm{mV}}
\global\long\def\gap{\mathrm{gap}}
\global\long\def\GB{\:\mathrm{GB}}
\global\long\def\ns{\:\mathrm{nS}}
\global\long\def\A{\mathbf{A}}

\global\long\def\A{{\text{A}^{-}}}

\global\long\def\a{\mathbf{a}}

\global\long\def\B{\mathbf{B}}

\global\long\def\b{\mathbf{b}}

\global\long\def\C{\mathbf{C}}

\global\long\def\c{\mathbf{c}}

\global\long\def\Ca{{\text{Ca}^{2+}}}

\global\long\def\cc{{^{*}}}

\global\long\def\Cl{{\text{Cl}^{-}}}

\global\long\def\Cov#1{\text{Cov}\left(#1\right)}

\global\long\def\d{\mathrm{d}}

\global\long\def\diff#1#2{{\displaystyle \frac{\text{d}#1}{\text{d}#2}}}

\global\long\def\D{\mathbf{D}}

\newcommandx\EW[2][usedefault, addprefix=\global, 1=]{\left\langle #2\right\rangle _{#1}}

\global\long\def\e{\mathbf{e}}

\global\long\def\erf{\mathrm{erf}}

\global\long\def\erfc{\mathrm{erfc}}

\global\long\def\Em{\mathbf{1}}

\global\long\def\Ex{\mathcal{E}}

\global\long\def\diag{\mathrm{diag}}

\global\long\def\E{\text{E}}

\global\long\def\ftr{\mathcal{F}}

\global\long\def\f{\mathbf{f}}

\global\long\def\Ftr#1#2{\mathfrak{F}[#1](#2)}

\global\long\def\iFtr#1#2{\mathfrak{F}^{-1}[#1](#2)}

\global\long\def\h{\mathbf{h}}

\global\long\def\G{\mathbf{G}}

\global\long\def\Hz{\:\mathrm{Hz}}

\global\long\def\In{\mathcal{I}}

\global\long\def\inp{\text{inp}}

\global\long\def\I{\text{I}}

\global\long\def\Int#1#2#3#4{\int\limits _{#3}^{#4} \text{d}#2\ #1}

\global\long\def\j{\mathbf{j}}

\global\long\def\J{\mathbf{J}}

\global\long\def\K{{\text{K}^{+}}}

\global\long\def\LN{\mathcal{L}_{0}}

\global\long\def\LO{\mathcal{L}_{1}}

\global\long\def\Lfp{L_{\mathrm{FP}}}

\global\long\def\M{\mathbf{M}}

\global\long\def\m{\mathbf{m}}

\global\long\def\mus{\:\mu\mathrm{s}}

\global\long\def\ms{\,\text{ms}}

\global\long\def\mV{\,\text{mV}}

\global\long\def\N{\mathbf{N}}

\global\long\def\n{\mathbf{n}}

\global\long\def\Na{{\text{Na}^{+}}}

\global\long\def\nuo{\nu_{0}}

\global\long\def\nr{n_{r}}

\global\long\def\OO{\mathbf{O}}

\global\long\def\P{\mathbf{P}}

\global\long\def\PV{P(V)}

\global\long\def\pF{\:\mathrm{pF}}

\global\long\def\Q{\mathbf{Q}}

\global\long\def\q{\mathbf{q}}

\global\long\def\R{\mathbf{R}}

\global\long\def\r{\mathbf{r}}

\global\long\def\res{\mathrm{Res}}

\global\long\def\rest{\text{rest}}

\global\long\def\s{\mathbf{s}}

\global\long\def\sps{\text{s}^{-1}}

\global\long\def\tilPV{\tilde{P}(V)}

\global\long\def\v{\mathbf{v}}

\global\long\def\V{\mathbf{V}}

\global\long\def\Var{\mathrm{Var}}

\global\long\def\Vth{V_{\theta}}

\global\long\def\Vr{V_{r}}

\global\long\def\x{\mathbf{x}}

\global\long\def\y{\mathbf{y}}

\global\long\def\T{\mathbf{T}}

\global\long\def\transp{\mathbf{^{\text{T}}}}

\global\long\def\uvec{\boldsymbol{u}}

\global\long\def\U{\mathbf{U}}

\global\long\def\Var#1{\text{Var}\left(#1\right)}

\global\long\def\vec#1{\boldsymbol{#1}}

\global\long\def\w{\boldsymbol{w}}

\global\long\def\W{\mathbf{W}}

\global\long\def\X{\mathbf{X}}

\global\long\def\y{\mathbf{y}}

\global\long\def\Y{\mathbf{Y}}

\global\long\def\taue{\tau_{e}}

\global\long\def\taum{\tau_{\mathrm{m}}}

\global\long\def\taus{\tau_{\mathrm{s}}}

\global\long\def\taur{\tau_{\mathrm{r}}}

\global\long\def\tauM{\tau_{\text{m}}}

\global\long\def\tauR{\tau_{\text{ref}}}

\global\long\def\nuext{\nu_{\mathrm{ext}}}

\global\long\def\defeq{\vcentcolon=}

\global\long\def\rdefeq{\mathrm{=\vcentcolon}}

\global\long\def\orderofmagnitude{\sim}

\global\long\def\unity{\mathds{1}}

\global\long\def\spikess{\;\mathrm{spikes}/\mathrm{s}}

\global\long\def\tav#1{\widehat{#1}}

\global\long\def\av#1{\overline{#1}}

\global\long\def\dotv{\mathord\cdot}

\noindent 

\noindent 

\noindent \begin{titlepage}\setcounter{page}{0}\pdfbookmark[1]{Title}{TitlePage}

\title{\textbf{\huge{}Integration of continuous-time dynamics in a spiking
neural network simulator}}
\maketitle
\begin{center}
\textbf{\large{}Jan Hahne$^{1\dagger*}$, David Dahmen$^{2\dagger}$,
Jannis Schuecker$^{2\dagger}$, Andreas Frommer$^{1}$, Matthias Bolten$^{3}$,
Moritz Helias$^{2,4}$ and Markus Diesmann$^{2,4,5}$}
\par\end{center}{\large \par}

\vspace{2cm}

\noindent$^{1}$\parbox[t]{12cm}{School of Mathematics and Natural
Sciences\\
Bergische Universit{\"a}t Wuppertal\\
Wuppertal, Germany}\\[3mm]

\noindent$^{2}$\parbox[t]{12cm}{Institute of Neuroscience and Medicine
(INM-6)\\
Institute for Advanced Simulation (IAS-6)\\
JARA BRAIN Institute I\\
J{\"u}lich Research Centre\\
J{\"u}lich, Germany}\\[3mm]

\noindent$^{3}$\parbox[t]{12cm}{Institut für Mathematik\\
Universität Kassel\\
Kassel, Germany}\\[3mm]

\noindent$^{4}$\parbox[t]{12cm}{Department of Physics\\
Faculty 1 \\
RWTH Aachen University\\
Aachen, Germany}\\[3mm]

\noindent$^{5}$\parbox[t]{12cm}{Department of Psychiatry, Psychotherapy
and Psychosomatics\\
Medical Faculty \\
RWTH Aachen University\\
Aachen, Germany}\\[3mm]

\vspace{0.5cm}\noindent$^{\dagger}$These authors contributed equally

\vspace{0.5cm}\noindent$^{*}$Correspondence to:\hspace{1em}\parbox[t]{11cm}{Jan
Hahne\\
Bergische Universit{\"a}t Wuppertal\\
42097 Wuppertal, Germany\\
\href{mailto:hahne@math.uni-wuppertal.de}{hahne@math.uni-wuppertal.de}

}\thispagestyle{empty}

\end{titlepage}
\begin{abstract}
Contemporary modeling approaches to the dynamics of neural networks
consider two main classes of models: biologically grounded spiking
neurons and functionally inspired rate-based units. The unified simulation
framework presented here supports the combination of the two for multi-scale
modeling approaches, the quantitative validation of mean-field approaches
by spiking network simulations, and an increase in reliability by
usage of the same simulation code and the same network model specifications
for both model classes. While most efficient spiking simulations rely
on the communication of discrete events, rate models require time-continuous
interactions between neurons. Exploiting the conceptual similarity
to the inclusion of gap junctions in spiking network simulations,
we arrive at a reference implementation of instantaneous and delayed
interactions between rate-based models in a spiking network simulator.
The separation of rate dynamics from the general connection and communication
infrastructure ensures flexibility of the framework. We further demonstrate
the broad applicability of the framework by considering various examples
from the literature ranging from random networks to neural field models.
The study provides the prerequisite for interactions between rate-based
and spiking models in a joint simulation. 
\end{abstract}

\section*{Keywords}

rate models, spiking neural network simulator, stochastic (delay)
differential equations, waveform relaxation, NEST

\section{Introduction\label{sec:Introduction}}

Over the past decades, multiple strategies of neural network modeling
have emerged in computational neuroscience. Functionally inspired
top-down approaches that aim to understand the computation in neural
networks typically describe neurons or neuronal populations in terms
of continuous variables, e.g. firing rates \citep{Hertz91,Schoener15}.
Rate-based models originate from the seminal works by \citet{Wilson72_1}
and \citet{Amari77} and were introduced as a coarse-grained description
of the overall activity of large-scale neuronal networks. Being amenable
to mathematical analyses and exhibiting rich dynamics such as multistability,
oscillations, traveling waves, and spatial patterns \citep{Coombes05,Roxin05,Bressloff12},
rate-based models have fostered progress in the understanding of memory,
sensory and motor processes \citep{Kilpatrick14}. Particular examples
include visuospatial working memory \citep{Camperi98_383,Laing02},
decision making \citep{Usher01,Bogacz06,Bogacz07}, perceptual rivalry
\citep{Laing02_39,Wilson03_14499}, geometric visual hallucination
patterns \citep{Ermentrout79a,Bressloff01_299}, ocular dominance
and orientation selectivity \citep{Ben-Yishai95,Bressloff03_1643,Stevens13_15747},
spatial navigation \citep{Samsonovich97_5900}, and movement preparation
\citep{Erlhagen02_545}. On the brain scale, rate models have been
used to study resting-state activity \citep{Deco11} and hierarchies
of time scales \citep{Chaudhuri2015_419}. Ideas from functional network
models have further inspired the field of artificial neuronal networks
in the domain of engineering \citep{Haykin08}.

In contrast, bottom-up approaches explicitly model individual neurons,
employing biophysically grounded spiking neuron models that simulate
the time points of action potentials. These models can explain a variety
of salient features of microscopic neural activity observed \emph{in
vivo}, such as spike-train irregularity \citep{Softky93,Vreeswijk96,Amit97,Shadlen98},
membrane-potential fluctuations \citep{Destexhe99_1531}, asynchronous
firing \citep{Brunel00,Ecker10,Renart10_587,Ostojic14}, correlations
in neural activity \citep{Gentet10_422,Okun2008_535,Helias13_023002},
self-sustained activity \citep{Ohbayashi03_233,Kriener2014_136},
rate distributions across neurons \citep{Griffith1966_516,Koch1989_292,Roxin11_16217}
and across laminar populations \citep{Potjans14_785}, as well as
resting state activity \citep{Deco12_3366}. 

Simulation of rate-based models goes back to the works by \citet{Grossberg73_213},
\citet{McClelland81}, \citet{Feldman82_205}, and the PDP group \citep{PDP86a}.
Various specialized tools have developed since then \citep{OReilly_simulators},
such as \textit{PDP++ \citep{McClelland89,OReilly00}, }the\textit{
Neural Simulation Language} \citep{Weitzenfeld02}, \textit{emergent
\citep{OReilly12}}, the \textit{MIIND} simulator \citep{deKamps_08},
the simulation platform \textit{DANA \citep{Rougier12_237}}, \textit{TheVirtualBrain
\citep{SanzLeon13}}, \textit{Topographica \citep{Bednar09_8}} and
the \textit{Neural Field Simulator \citep{Nichols15}}. Similarly,
efficient simulators for spiking neural networks have evolved with
different foci ranging from detailed neuron morphology (NEURON \citep{Carnevale06},
GENESIS \citep{Bower07_1383}) to an abstraction of neurons to a single
point in space (NEST \citep{Nest2100}, BRIAN \citep{Goodman13},
\citealp[see][for a review]{Brette07_349}). Such open-source software
supports maintainability, reproducibility, and exchangeability of
models and code, as well as community driven development. However,
these tools are restricted to either rate-based or spike-based models
only.

The situation underlines that bottom-up and top-down strategies are
still mostly disjoint and a major challenge in neuroscience is to
form a bridge between the spike- and rate-based models \citep{Abbott16_350},
and, more generally, between the fields of computational neuroscience
and cognitive science. From a practical point of view, a common simulation
framework would allow the exchange and the combination of concepts
and code between the two descriptions and trigger interaction between
the corresponding communities. This is in particular important since
recent advances in simulation \citep{Djurfeldt08_31,Hines08_203,Hines10_1,Hines11_49,Helias12_26,Kunkel14_78}
and computing technology \citep{stephan2015juqueen}  enable full-density
bottom-up models of complete circuits \citep{Potjans14_785,Markram2015_456}.
In particular, it has become feasible to build spiking models \citep{Schmidt16_arxiv_v3}
that describe the same macroscopic system as the rate-based descriptions
\citep{Chaudhuri2015_419}. 

Mean-field theories pave the way to relate spiking and rate-based
descriptions of neuronal dynamics \citep{Deco-2008_e1000092}. Population
density methods using Fokker-Planck theory can be used to determine
stationary state activities of spiking networks \citep{Siegert51,Brunel00}.
The dynamics of rate fluctuations around this background activity
can be obtained using linear response theory \citep{Brunel99,Lindner05_061919,Ostojic11_e1001056,Trousdale12_e1002408,Grytskyy13_131,Schuecker15_transferfunction},
moment expansions for mode decompositions of the Fokker-Planck operator
\citep{Mattia02_051917,Mattia04_052903,Deco-2008_e1000092}, or other
specific ansatzes \citep{Montbrio15}. An alternative derivation of
rate-based dynamics aims at a closure of equations for synaptic currents
of spiking networks in a coarse-graining limit by replacing spiking
input with the instantaneous firing rate \citep{Bressloff12}. Using
field-theoretical methods \citep{Buice13_1} that were originally
developed for Markovian network dynamics \citep{Buice07_051919,Buice09_377}
allows a generalization of this approach to fluctuations in the input
\citep{Bressloff15}. In any case, the cascade of simplifications
from the original spiking network to the rate-based model involves
a combination of approximations which are routinely benchmarked in
comparative simulations of the two models. A unified code base that
features both models would highly simplify these benchmarks rendering
duplication of code obsolete.

Rate neurons typically represent populations of spiking neurons. Thus,
a hybrid model, employing both types of neuron models in a multi-scale
modeling approach, would contain a relatively large number of spiking
neurons compared to the number of rate units. Furthermore a downscaling
of a spiking network cannot be performed without changing the dynamics
\citep{Albada15} and thus it is crucial that a common simulation
framework is able to handle real-sized spiking networks. In addition,
the employed mean-field theories exploit the large number of neurons
in biological networks. In fact, they are strictly valid only in the
thermodynamic limit $N\rightarrow\infty$ \citep{Helias14}. Therefore,
in the above mentioned benchmarks, the spiking networks are typically
large. Thus, a common simulation framework should be optimized for
spiking neurons rather than rate-based models.  

Current spiking network simulators solve the neuronal dynamics in
a distributed and parallel manner. They exploit the point-event like
nature of the spike interaction between neurons, for example in event-based
simulation schemes. The latter, however, can not be employed in the
context of rate-based models which require continuous interactions
between units. Spiking point-neuron models furthermore interact in
a delayed fashion. The delays mimic the synaptic transmission and
the propagation times along axons and dendrites. For the duration
of the minimal delay $d_{\mathrm{min}}$ in a network, the dynamics
of all neurons is decoupled. Hence, during $d_{\mathrm{min}}$, the
neurons can be updated independently without requiring information
from other neurons. Distributed processes therefore need to communicate
spikes only after this period \citep{Morrison05a}. Due to considerable
latencies associated with each communication, this scheme significantly
improves performance and scalability of current simulators. In contrast,
rate based-models \citep[see][and references therein]{Bressloff12}
consider instantaneous interactions between neurons. A priori, this
requires communication of continuous state variables between neurons
at each time step. 

The present study provides the concepts and a reference implementation
for the embedding of continuous-time dynamics in a spiking network
simulator. In order to exploit existing functionality we choose as
a platform the open source simulation code NEST \citep{Gewaltig_07_11204,Nest2100}
which is scalable software that can be used on machines ranging from
laptops to supercomputers. The software is utilized by a considerable
user community and equipped with a Python interface, support for the
construction of complex networks, and mechanisms to shield the neuroscientist
from the difficulties of handling a model description, potentially
including stochastic components, in a distributed setting \citep{Morrison05a,Plesser15_1849}.
Within this framework we introduce an iterative numerical solution
scheme that reduces communication between compute nodes. The scheme
builds on the waveform-relaxation technique \citep{Lelarasmee82}
already employed for gap-junction interactions \citep{Hahne15_00022}.

Our study begins with a brief review of numerical solution schemes
for ordinary and stochastic (delay) differential equations in \prettyref{subsec:Stochastic-differential-equation}
and their application to neural networks in \prettyref{subsec:Rate-models}.
Subsequently, we develop the concepts for embedding rate-based network
models into a simulation code for spiking networks, adapt the waveform-relaxation
scheme, and detail an extendable implementation framework for neuron
models in terms of \texttt{C++} templates \prettyref{subsec:NEST-implementation}.
In \prettyref{sec:Results}, different numerical schemes are evaluated
as well as the scalability of our reference implementation. We illustrate
the applicability of the framework to a broad class of network models
on the examples of a linear network model \citep{Grytskyy13_131},
a nonlinear network model \citep{Sompolinsky88_259,Goedeke16_arxiv},
a neural field model \citep{Roxin05}, and a mean-field description
\citep{Wong_06} of the stationary activity in a model of the cortical
microcircuit \citep{Potjans14_785,Schuecker16_arxiv_v3}. Straight-forward
generalizations are briefly mentioned at the end of the Results section,
before the work concludes with the Discussion in \prettyref{sec:Discussion}.
The technology described in the present article will be made available
with one of the next major releases of the simulation software NEST
as open source. The conceptual and algorithmic work is a module in
our long-term collaborative project to provide the technology for
neural systems simulations \citep{Gewaltig_07_11204}.

\section{Material and methods\label{sec:Material-and-methods}\label{subsec:Stochastic-differential-equation}}

Rate-based single neuron and population models are described in terms
of differential equations that often include delays and stochastic
elements. Before we turn to the implementation of such models in computer
code (\prettyref{subsec:NEST-implementation}) we review how such
systems are mathematically solved and in particular how the stochastic
elements are commonly interpreted with the aim to avoid an ad hoc
design. A stochastic differential equation (SDE) is defined by the
corresponding stochastic integral equation. Let $W(t)$ denote a Wiener
process, also called Standard Brownian motion. For the initial condition
$X(t_{0})=X_{0}$ an It\^{o}-SDE in its most general form satisfies

\begin{equation}
X(t)=X_{0}+\int_{t_{0}}^{t}a(s,X(s))\,ds+\int_{t_{0}}^{t}b(s,X(s))\,dW(s)\,,\label{eq:general-sde-integral}
\end{equation}

\noindent where the second integral is an It\^{o} integral

\[
\int_{t_{0}}^{t}Y(s)\,dW(s):=\lim_{n\to\infty}\sum_{i=1}^{n}Y_{i-1}\cdot(W_{i}-W_{i-1})
\]

\noindent with $Y_{i}=Y(t_{0}+i\cdot\frac{t-t_{0}}{n})$ and $W_{i}=W(t_{0}+i\cdot\frac{t-t_{0}}{n})$.
Alternatively, the second integral can be chosen as a Stratonovich
integral, indicated by the symbol $\circ$,

\[
\int_{t_{0}}^{t}Y(s)\circ dW(s):=\lim_{n\to\infty}\sum_{i=1}^{n}\frac{Y_{i-1}+Y_{i}}{2}\,(W_{i}-W_{i-1})
\]

\noindent which approximates $Y(s)$ with the mid-point rule. In this
case, the corresponding SDE is called a Stratonovich-SDE. We refer
to \citet{Kloeden92} and \citet{Gardiner04} for a derivation and
a deeper discussion on the differences between the two types of stochastic
integrals. In the case of additive noise ($b(t,X(t))=b(t)$) the It\^{o}
and Stratonovich integrals coincide. If furthermore the noise is
constant ($b(t,X(t))=\sigma=\mathrm{const.}$) the integrals can be
solved analytically

\[
\int_{t_{0}}^{t}\sigma\,dW(s)=\int_{t_{0}}^{t}\sigma\circ dW(s)=\lim_{n\to\infty}\sigma\cdot\sum_{i=1}^{n}(W_{i}-W_{i-1})=\sigma\cdot(W(t)-W(t_{0}))
\]

\noindent with $W(t)-W(t_{0})\sim\mathcal{N}(0,t-t_{0})$. In the
following, we focus on It\^{o}-SDEs only.

\noindent The differential notation corresponding to \prettyref{eq:general-sde-integral}
reads

\begin{equation}
dX(t)=a(t,X(t))\,dt+b(t,X(t))\,dW(t)\label{eq:general-sde}
\end{equation}

\noindent and denotes an informal way of expressing the integral equation.
Another widely used differential notation, called the Langevin form
of the SDE, is mostly employed in physics. It reads

\begin{equation}
\frac{dX(t)}{dt}=a(t,X(t))+b(t,X(t))\,\xi(t)\,,\label{eq:sde-langevin}
\end{equation}

\noindent where $\xi(t)$ is a Gaussian white noise with $\langle\xi(t)\rangle=0$
and $\langle\xi(t)\xi(t^{\prime})\rangle=\delta(t-t^{\prime})$. Using
the Fokker-Planck equation one obtains

\[
\int_{0}^{t}\xi(t^{\prime})dt^{\prime}=W(t)\,,
\]

\noindent which is a paradox, as one can also show that $W(t)$ is
not differentiable \citep[Chapter 4]{Gardiner04}. Mathematically
speaking this means that \prettyref{eq:sde-langevin} is not strictly
well-defined. The corresponding stochastic integral equation 

\[
X(t)=X_{0}+\int_{t_{0}}^{t}a(s,X(s))\,ds+\int_{t_{0}}^{t}b(s,X(s))\,\xi(s)\,ds\,,
\]

\noindent however, can be interpreted consistently with \prettyref{eq:general-sde-integral}
as $dW(t)\equiv\xi(t)dt$.

\subsection{Approximate numerical solution of SDEs\label{subsec:Approximate-numerical-solution}}

Similar to ordinary differential equations most stochastic differential
equations cannot be solved analytically. Neuroscience therefore relies
on approximate numerical schemes to obtain the solution of a given
SDE. This section presents some basic numerical methods. Let $\Delta t$
denote the fixed step size, $t_{k}=t_{0}+k\Delta t$ the grid points
of the discretization for $k=0,\ldots,n$, and $X_{k}$ the approximation
for $X(t_{k})$ obtained by the numerical method, at which $X_{0}$
is the given initial value. We consider systems of $N$ stochastic
differential equations

\begin{equation}
dX(t)=a(t,X(t))\,dt+b(t,X(t))\,dW(t)\label{eq:system-sdes}
\end{equation}

\noindent with initial condition $X(t_{0})=X_{0}$. Here, $X(t)=(X^{1}(t),\ldots,X^{N}(t))$
and $W(t)=(W^{1}(t),\ldots,W^{N}(t))$ denote $N$-dimensional vectors
and $a:\mathbb{R}^{N}\rightarrow\mathbb{R}^{N}$ and $b:\mathbb{R}^{N}\rightarrow\mathbb{R}^{N}$
are $N$-dimensional functions. $W(t)$ is an $N$-dimensional Wiener
process, i.e., the components $W^{i}(t)$ are independent and identically
distributed.

\paragraph{Euler-Maruyama}

The Euler-Maruyama method is a generalization of the forward Euler
method for ordinary differential equations (ODE). Accordingly, it
approximates the integrands in \prettyref{eq:general-sde-integral}
with their left-sided values. The update formula reads

\begin{equation}
X_{k+1}=X_{k}+a(t_{k},X_{k})\cdot\Delta t+b(t_{k},X_{k})\cdot\Delta W_{k}\label{eq:euler-mayurana}
\end{equation}
with $\Delta W_{k}=W(t_{k+1})-W(t_{k})\sim\mathcal{N}(0,\Delta t)$
for $k=0,\ldots,n-1$.

\paragraph{Semi-implicit Euler}

The (semi-)implicit Euler method is a generalization of the backwards
Euler method for ODEs. The update formula reads

\begin{equation}
X_{k+1}=X_{k}+a(t_{k+1},X_{k+1})\cdot\Delta t+b(t_{k},X_{k})\cdot\Delta W_{k}\label{eq:semi-implicit-euler}
\end{equation}

\noindent with $\Delta W_{k}\sim\mathcal{N}(0,\Delta t)$ for $k=0,\ldots,n-1$.
The resulting scheme requires the solution of a system of nonlinear
algebraic equations. Standard techniques for the solution of the system
are Newton iteration and fixed-point iteration \citep{Kelley95}.
The method is sometimes called semi-implicit, because the function
$b$ is still evaluated at $(t_{k},X_{k})$ instead of $(t_{k+1},X_{k+1})$.
However, a fully implicit Euler scheme for SDEs is not practicable
(see \citet{Kloeden92}, Chapter $9.8$) and thus the term implicit
Euler usually refers to the semi-implicit method.

\paragraph{Exponential Euler}

The exponential Euler method relies on the assumption that $a(t,X(t))$
consists of a linear part and a nonlinear remainder, i.e.,

\[
a(t,X(t))=A\cdot X(t)+f(t,X(t))
\]

\noindent with $A\in\mathbb{R}^{N\times N}$. The idea is to solve
the linear part exactly and to approximate the integral of the nonlinear
remainder and the It\^{o} integral with an Euler-like approach. Variation
of constants for \prettyref{eq:system-sdes} yields

\[
X(t)=e^{A(t-t_{0})}X_{0}+\int_{t_{0}}^{t}e^{A(t-s)}f(s,X(s))\,ds+\int_{t_{0}}^{t}e^{A(t-s)}b(s,X(s))\,dW(s)\,.
\]

\noindent There are several versions of stochastic exponential Euler
methods that differ in the approximation of the integral. Unfortunately
a standardized nomenclature to distinguish the methods is so far missing.
The simplest approach, sometimes named stochastic Lawson-Euler scheme
\citep[e.g. in][]{Komori14}, approximates the integrands with their
left-sided values:

\[
X_{k+1}=e^{A\Delta t}X_{k}+e^{A\Delta t}f(t_{k},X_{k})\cdot\Delta t+e^{A\Delta t}b(t_{k},X_{k})\cdot\Delta W_{k}\,.
\]

\noindent More advanced schemes approximate the nonlinear part by
keeping $f(s,X(s))$ constant for $[t_{0},t)$ and solving the remaining
integral analytically

\[
\int_{t_{0}}^{t}e^{A(t-s)}f(s,X(s))\,ds\approx\int_{t_{0}}^{t}e^{A(t-s)}f(t_{0},X(t_{0}))\,ds=A^{-1}(e^{A(t-t_{0})}-I)\cdot f(t_{0},X(t_{0}))\,.
\]
Here $I$ denotes the $N\times N$ identity matrix. The same technique
can be used for the It\^{o} integral

\begin{equation}
\int_{t_{0}}^{t}e^{A(t-s)}b(s,X(s))\,dW(s)\approx\int_{t_{0}}^{t}e^{A(t-s)}b(t_{0},X(t_{0}))\,dW(s)\,.\label{eq:ito-approximation}
\end{equation}

\noindent For a single SDE, \citet{Shoji11} proposed a method where
the remaining integral $\int_{t_{0}}^{t}e^{a(t-s)}\,dW(s)$ with $a\in\mathbb{R}$
is approximated by $\int_{t_{0}}^{t}\alpha\,dW(s)$, such that $\alpha\in\mathbb{R}$
is chosen to minimize the mean-square error. This results in a similar
approximation as for the nonlinear part. \citet{Komori14} adapted
this approach for systems of SDEs. The scheme reads

\[
X_{k+1}=e^{A\Delta t}X_{k}+A^{-1}(e^{A\Delta t}-I)\cdot f(t_{k},X(t_{k}))+\frac{1}{\Delta t}\cdot A^{-1}(e^{A\Delta t}-I)\cdot b(t_{k},X_{k})\cdot\Delta W_{k}\,.
\]
Alternatively, calculating the variance of $X(t)$ within the approximation
\prettyref{eq:ito-approximation}, amounts to \citep{Adamu11}

\[
\Var{X(t)}=b(t_{0},X(t_{0}))^{2}\cdot\Var{\int_{t_{0}}^{t}e^{A(t-s)}\,dW(s)}=b(t_{0},X(t_{0}))^{2}\cdot A^{-1}\left(\frac{e^{2A(t-t_{0})}-I}{2}\right)\,.
\]
The corresponding scheme reads

\begin{equation}
X_{k+1}=e^{A\Delta t}X_{k}+A^{-1}(e^{A\Delta t}-I)\cdot f(t_{k},X(t_{k}))+\sqrt{A^{-1}\left(\frac{e^{2A\Delta t}-I}{2}\right)}\cdot b(t_{k},X_{k})\cdot\eta_{k}\label{eq:stochastic-exp-euler}
\end{equation}
with $\eta_{k}\sim\mathcal{N}(0,1)$ and yields the exact solution
of the system if $a(t,X(t))=A\cdot X(t)$ and $b(t,X(t))=\text{const.}$,
since $X(t)$ has Gaussian statistics in this case \citep{Risken96}.
Therefore in the following we exclusively employ \prettyref{eq:stochastic-exp-euler}
and just refer to it as the stochastic exponential Euler scheme. For
more detailed reviews on the different stochastic exponential Euler
methods we refer to \citep{Adamu11} and \citep{Komori14}.

\subsection{Network of rate models\label{subsec:Rate-models}}

We here consider networks of $N$ rate-based model neurons where each
neuron receives recurrent input from the network. The system fulfills
the It\^{o}-SDEs

\begin{equation}
\tau^{i}dX^{i}(t)=\left[-X^{i}(t)+\mu^{i}+\phi\left(\sum_{j=1}^{N}w^{ij}\psi\left(X^{j}(t-d^{ij})\right)\right)\right]\,dt+\sqrt{\tau^{i}}\sigma^{i}\,dW^{i}(t)\quad\quad i=1,\ldots,N\label{eq:network-OUP}
\end{equation}
with possibly nonlinear input-functions $\phi(x)$ and $\psi(x)$,
connection weights $w^{ij}$, mean input $\mu^{i}$, and optional
delays $d^{ij}\geq0$. The corresponding Fokker-Planck equation shows
that the parameter $\sigma^{i}\geq0$ controls the variance of $X^{i}(t)$
and the time constant $\tau^{i}>0$ its temporal evolution. For readability,
from here on we omit neuron indices for $\sigma,\tau,\mu$, and $d$.
The considered class of rate models only contains additive noise.
Therefore, as noted above, the system \prettyref{eq:network-OUP}
can be written as Stratonovich-SDEs without the need for change in
the employed numerical methods. For an illustrative purpose we explicitly
state the different explicit solution schemes for the network dynamics
\prettyref{eq:network-OUP} with $d=0$. The Euler-Maruyama update
step reads 

\begin{equation}
X_{k+1}^{i}=X_{k}^{i}+\left[-X_{k}^{i}+\mu+\phi\left(\sum_{j=1}^{N}w^{ij}\psi\left(X_{k}^{j}\right)\right)\right]\,\frac{1}{\tau}\Delta t+\frac{1}{\sqrt{\tau}}\sigma\Delta W_{k}^{i}\:.\label{eq:network_euler_maruyama}
\end{equation}
For nonlinear $\phi(x)$ or $\psi(x)$ the exponential Euler update
step is 

\begin{equation}
X_{k+1}^{i}=e^{-\Delta t/\tau}X_{k}^{i}+\left(1-e^{-\Delta t/\tau}\right)\left[\mu+\phi\left(\sum_{j=1}^{N}w^{ij}\psi\left(X_{k}^{j}\right)\right)\right]+\sqrt{\dfrac{1}{2}(1-e^{-2\Delta t/\tau})}\,\sigma\eta_{k}^{i}\,\label{eq:network_exponential_euler}
\end{equation}
with $\eta_{k}^{i}\sim\mathcal{N}(0,1)$. As $A=-I$ is a diagonal
matrix, the exponential Euler scheme does not rely on a matrix exponential,
but decomposes into $N$ equations with scalar exponential functions.
Note that with a linear choice, $\phi(x)=\psi(x)=x$, the system of
SDEs can be written in matrix notation 

\begin{equation}
\tau dX(t)=\left[A\cdot X(t)+\mu\right]\,dt+\sqrt{\tau}\sigma dW(t)\quad\quad i=1,\ldots,N\label{eq:linear-network-OUP}
\end{equation}

\noindent with $A=-I+W$ and $W=(w^{ij})_{N\times N}$. Here the stochastic
exponential Euler scheme \prettyref{eq:stochastic-exp-euler} yields
the exact solution of the system.

\noindent The numerical schemes presented in \prettyref{subsec:Approximate-numerical-solution}
are developed for SDEs ($d=0$), but can analogously be used for stochastic
delay differential equations (SDDEs) ($d>0$), if the delay $d$ is
a multiple of the step size $\Delta t$. For the calculation of the
approximation $X_{k+1}^{i}$ in time step $k+1$ the recurrent input
is then evaluated from $X_{k-\frac{d}{\Delta t}}^{j}$, i.e. from
$\frac{d}{\Delta t}$ steps earlier.

\subsection{Implementation in spiking network simulation code\label{subsec:NEST-implementation}}

This section describes the embedding of rate-based models (\prettyref{subsec:Rate-models})
in a simulation code for spiking neuronal networks. Examples how
to create, connect and record activity from rate models will be made
available with the release of our reference implementation of the
continuous-time dynamics in the simulation code NEST.

The software architecture for rate models is based on existing concepts:
\citet{Morrison05a} describe distributed buffers for the storage
of delayed interactions and the technique to consistently generate
random numbers in a distributed setting, and \citet{Hahne15_00022}
introduce so called \texttt{SecondaryEvent}s, that allow the communication
of any kind of data between pairs of neurons. These components are
designed to be compatible with the parallel and distributed operation
of a simulation kernel for spiking neuronal networks, ensuring an
efficient use of clusters and supercomputers \citep{Helias12_26}.
This allows researchers to easily scale up network sizes to more realistic
number of neurons. The highly parallelizable structure of modern simulation
codes for spiking neuronal networks, however, also poses restrictions
on the utilizable numerical methods.

\subsubsection{Restrictions\label{par:Restrictions}}

Parallelization for spiking neuronal networks is achieved by distributing
neurons over compute nodes. Since the dynamics of spiking neurons
(in the absence of gap junctions) is decoupled for the duration of
the minimal synaptic delay $d_{\textrm{min}}$ of the connections
in the network, the states of the neurons can be propagated independently
for this time interval. Thus it is sufficient to specify solvers on
the single-neuron level. The spike times, i.e. the mediators of interaction
between neurons, are then communicated in steps of $d_{\textrm{min}}$.
In contrast, in the case of interactions via gap junctions \citep{Hahne15_00022}
or rates, the single-neuron dynamics depends on continuous state variables,
membrane potential or rate, of other neurons. These continuous variables
need to be communicated and the mechanism for this is the \texttt{SecondaryEvent}
introduced in the gap-junction framework by \citet{Hahne15_00022}. 

Furthermore, the global connectivity of the network is unknown to
the single neuron. The neuron object sends and receives events handled
by the network manager on the compute node harboring the neuron. However,
the network manager only knows the incoming connections of the neurons
on the compute node.

This structure makes it impossible to employ the implicit Euler scheme
\prettyref{eq:semi-implicit-euler} with Newton iteration, which would
require the simultaneous solution of a system of nonlinear algebraic
equations with information distributed over all compute nodes. It
is however possible to use implicit schemes with fixed-point iteration.
To this end, the corresponding scheme needs to be formulated as a
fixed-point iteration on the single-neuron level and the updated influences
of other neurons have to be communicated in every iteration. The gap
junction framework by \citet{Hahne15_00022} already specifies an
iterative method to advance the state of the network by one time step
with global accuracy control. Therefore, we investigate in \prettyref{subsec:Comparison-of-numerical}
if for rate-based network models the payoff of an implicit scheme
is large enough to justify the additional effort of an iterative solver.

The restricted knowledge of connectivity also limits the usage of
the exponential Euler method. In the case of a linear rate model,
we are not able to add the influence from all other rate neurons to
the matrix $A$ in \prettyref{eq:linear-network-OUP}, because most
of these connections are unknown at the single-neuron level. Therefore,
we use the exponential Euler method with $A=-I$ resulting in the
update formula \prettyref{eq:network_exponential_euler}. This also
has the benefit of avoiding the need to numerically evaluate a general
matrix exponential as $A$ is a diagonal matrix (see \prettyref{subsec:Rate-models}
for details).

\subsubsection{Implementation\label{par:Implementation-details}}

\noindent This section describes the additional data structure required
for the implementation of rate-based models. As a result of the previous
section and our analysis of the numerical schemes in \prettyref{sec:Results}
we restrict the discussion to the exponential Euler method where we
assume $A=-I$ and identify $\Delta t=h$ with $h$ denoting the global
computation step size \citep{Morrison05a}. We have to distinguish
the cases of connections with delay ($d>0$) and connections without
delay ($d=0$). The former case is similar to spiking interaction:
assuming a connection from neuron $i$ to neuron $j$, the rate of
neuron $i$ needs to be available at neuron $j$ after $\frac{d}{h}$
additional time steps. This can be ensured if the delay of the connection
is considered in the calculation of the minimal delay $d_{\min}$
that determines the communication interval. After communication the
rate values are stored in a ring buffer of neuron $j$ until they
are due \citep{Morrison08_267}. In the case of an instantaneous connection,
the rate of neuron $i$ at time $t_{0}$ needs to be known at time
$t_{0}$ at the process which updates neuron $j$ from $t_{0}$ to
$t_{0}+h$. Therefore, communication in every step is required for
instantaneous rate connections, i.e. setting $d_{\min}=h$.

Due to the conceptual differences between instantaneous and delayed
interactions \citep[for the conceptual difference in the case of spiking interaction see ][]{Morrison08_267}
we define two different connection types and associated events: The
connection type for connections with delay is called \texttt{delay\_rate\_connection}
and is associated with the new \texttt{SecondaryEvent} type \texttt{DelayRateNeuronEvent}.
Connections without delay are implemented with the connection type
\texttt{rate\_connection} with the corresponding secondary event \texttt{RateNeuronEvent}.

The template class \texttt{rate\_neuron\_ipn} provides a base implementation
for rate models of category \prettyref{eq:network-OUP}. In this implementation
neurons can handle both \texttt{DelayRateNeuronEvent} and \texttt{RateNeuronEvent}
allowing for simultaneous use of instantaneous and delayed connections.
To represent the nonlinearites $\phi(x)$ and $\psi(x)$ the class
contains an object \texttt{gain\_} the type of which is determined
by the template parameter \texttt{TGainfunction}. The ending \texttt{ipn}
indicates input noise, as the noise directly enters the r.h.s. of
\eqref{eq:network-OUP}. A constant boolean class member \texttt{linear\_summation}\_
of \texttt{rate\_neuron\_ipn} determines if the nonlinearity expressed
by the \texttt{operator()} of the object \texttt{gain\_} should be
interpreted as $\phi(x)$ (true) or $\psi(x)$ (false). The respective
other function is assumed to be the identity function and the default
setting for \texttt{linear\_summation}\_ is true. While to our knowledge
this implementation covers the majority of neuron models, the evaluation
of the boolean parameter \texttt{linear\_summation}\_ in every update
step of each neuron could be improved in terms of efficiency if the
type of nonlinearity would be decided upon at compile time. In the
present architecture this would, however, result in twice as many
template instances for a given set of gain functions. With the future
capabilities of code generation \citep{Plotnikov_2016} in mind it
might be beneficial to elevate the constant boolean member object
to a constant template parameter to allow compilers efficient preprocessing
and at the same time profit from the code reliability achievable by
modern C++ syntax. The present base implementation reduces the effort
of creating a specific rate model of category \prettyref{eq:network-OUP}
to the specification of an instance of the template class \texttt{TGainfunction}.
Afterwards the actual neuron model can be defined with a simple typedef
like e.g.
\noindent \begin{center}
\texttt{typedef rate\_neuron\_ipn< nest::gainfunction\_lin\_rate >
lin\_rate\_ipn;}
\begin{table}[t]
\begin{centering}
\begin{tabular}{|c|c|}
\hline 
gain model & $\phi(x)$ or $\psi(x)$\tabularnewline
\hline 
\hline 
\texttt{lin\_rate} & $x$\tabularnewline
\hline 
\texttt{tanh\_rate} & $\tanh(g\cdot x)$ with $g\in\mathbb{R}$\tabularnewline
\hline 
\texttt{thresholdlin\_rate} & $g\cdot(x-\theta)\cdot H(x-\theta)$ with $g\in\mathbb{R}$ \tabularnewline
\hline 
\end{tabular}
\par\end{centering}
\caption{\textbf{Template-derived rate-based neuron models.} The table shows
the gain functions of the rate-based neuron models available in the
NEST reference implementation. The name of a particular neuron model
is formed by \texttt{\textsl{<}}\texttt{gain model>\_ipn}.\label{tab:Implemented-rate-models}}
\end{table}
\par\end{center}

\prettyref{tab:Implemented-rate-models} gives an overview of rate
models of the NEST reference implementation. These models serve as
a reference for the implementation of customized neuron models. Activity
of rate neurons can be recorded using the \texttt{multimeter} and
the recordable \texttt{rate.}

\subsubsection{Waveform-relaxation techniques\label{par:Waveform-relaxation-techniques}}

\noindent The instantaneous connections between rate-based models
requires communication in every time step, which impairs the performance
and scalability. On supercomputers communication is particularly expensive,
because it is associated with a considerable latency. Therefore, we
also study an alternative iterative approach based on waveform-relaxation
techniques that allows us to use communication on a coarser time grid.
As outlined above, in a simulator for spiking neuronal networks the
communication intervals are defined by the minimal delay $d_{\mathrm{min}}$
in the network. For simulations with instantaneous connections only,
we attempt to reduce the communication load by setting the minimal
delay to an arbitrary user specified value given by the parameter
\texttt{wfr\_comm\_interval} (see \prettyref{tab:wfr-parameters}).
In case of additional delayed connections, the actual communication
interval for waveform relaxation then follows as $min\left(d_{\min},\texttt{wfr\_comm\_interval}\right)$.
For details on waveform-relaxation methods and their application in
the neuroscience context we refer to \citet{Hahne15_00022}. Originally
these methods were developed \citep{Lelarasmee82} and investigated
\citep[see e.g.][]{Miekkala87} for ODEs. More recently waveform relaxation
methods have also been analyzed for SDEs \citep{Schurz05} and successfully
applied to large systems of SDEs \citep{Fan13}.

\begin{figure}[!tbh]
\begin{centering}
\includegraphics{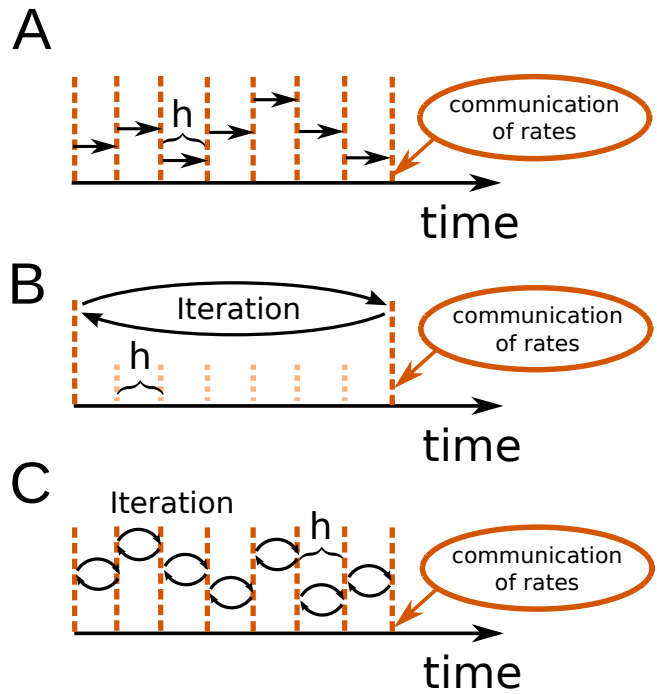}
\par\end{centering}
\caption{\textbf{Different communication strategies for distributed simulations.}
Distance between neighboring dotted orange lines indicates computation
time step of size $h$. Distance between neighboring dashed red lines
symbolize one communication interval where rates (and other events
like spike events) are communicated at the end of the interval. \textbf{(A)}
Straight-forward solution for rate-based models: rates are communicated
in every time step. \textbf{(B)} Iterative approach using waveform
relaxation: rates are communicated only after $\frac{d_{\min}}{h}$
steps and the entire interval is solved repeatedly. \textbf{(C)} Iterative
approach with communication in every step.\label{fig:com_strategies}}
\end{figure}

\noindent \prettyref{fig:com_strategies}B illustrates the concept
of the iterative approach in contrast to the standard procedure in
panel A. The iterative approach requires the repeated solution of
all time steps in the communication interval and converges to the
solution obtained with the standard approach (\prettyref{fig:com_strategies}A).
The iteration terminates when a user chosen convergence tolerance
\texttt{wfr\_tol} (see \prettyref{tab:wfr-parameters}) is met. If
the method needs less than $d_{min}/h$ iterations, the approach reduces
the overall number of communications required to obtain the solution.
In conclusion, the avoidance of communication in every step comes
for the price of additional computational load.

The coupling of neurons via gap junctions is instantaneous and continuous
in time and thus constitutes a very similar problem to the rate dynamics.
In order to combine gap junctions with spiking dynamics \citet{Hahne15_00022}
already devised an iterative technique. The dynamics of a neuron model
supporting gap junctions is solved with an adaptive step-size ODE-solver,
routinely carrying out several steps of the employed numerical method
within one global computation time step $h$. The communication of
a cubic interpolation of the membrane potential provides the solver
with additional information, resulting in a more accurate solution
than the one obtained from the standard approach. For rate-based models
this approach is however impossible. The combination of an iterative
method with an adaptive step-size solver is not applicable to SDEs,
where the noise in each time step constitutes a random number. However,
an iterative approach with fixed step size $\Delta t=h$ is applicable,
as long as we ensure that the random noise applied to the neurons
remains the same in every iteration. In \prettyref{subsec:Performance}
we investigate the performance of the iterative (\prettyref{fig:com_strategies}B)
and the standard approach (\prettyref{fig:com_strategies}A) with
a focus on large network simulations on supercomputers. In our reference
implementation waveform relaxation can be enabled or disabled by a
parameter \texttt{use\_wfr}. Note that in the traditional communication
scheme for spiking neuronal networks \citep{Morrison05a} the first
communication occurs earliest at the end of the first update step.
Therefore, in the absence of waveform relaxation, the initial input
to neurons from the network is omitted.

\prettyref{fig:com_strategies}C shows an alternative iterative approach
also feasible within our framework. While this scheme is not needed
for the exponential Euler method investigated in the present work,
it can be employed to perform a fixed point iteration in order to
obtain the solution of the implicit Euler method. However, \prettyref{subsec:Comparison-of-numerical}
demonstrates that this is not an efficient option for the integration
of rate-based models in distributed simulations. 

\begin{table}[t]
\begin{centering}
\begin{tabular}{|l|l|l|>{\raggedright}m{0.5\textwidth}|}
\hline 
parameter name & type & default & \multirow{1}{0.5\textwidth}{description}\tabularnewline
\hline 
\hline 
\texttt{use\_wfr} & \texttt{bool} & \texttt{true} & Boolean parameter to enable (\texttt{true}) or disable (\texttt{false})
the use of the waveform relaxation technique. If disabled and any
rate-based neurons (or neurons supporting gap junctions) are present,
communication in every step is automatically activated ($d_{\min}=h$).\tabularnewline
\hline 
\texttt{wfr\_comm\_interval} & \texttt{double} & $1.0\ms$ & Instantaneous rate connections (and gap junctions) contribute to the
calculation of the minimal network delay with $min\left(d_{\min},\texttt{wfr\_comm\_interval}\right)$.
This way the length of the iteration interval of the waveform relaxation
can be regulated.\tabularnewline
\hline 
\texttt{wfr\_tol} & \texttt{double}  & $10^{-4}$ & Convergence criterion for waveform relaxation. The iteration is stopped
if the rates of all neurons change less than \texttt{wfr\_tol} from
one iteration to the next.\tabularnewline
\hline 
\texttt{wfr\_max\_iterations} & \texttt{int} & $15$ & Maximum number of iterations performed in one application of the waveform
relaxation. If the maximum number of iterations has been carried out
without reaching the accuracy goal the algorithm advances system time
and the reference implementation issues a warning.

Additional speed-up in the simulation of rate-based neurons can only
be achieved by \texttt{$\texttt{wfr\_max\_iterations}<d_{min}/h$.}\tabularnewline
\hline 
\texttt{wfr\_interpolation\_order} & \texttt{int} & $3$ & This parameter is exclusively used for gap junctions \citep[see][Sec. 2.1.2]{Hahne15_00022}
and has no influence on the simulation of rate-based models.\tabularnewline
\hline 
\end{tabular}
\par\end{centering}
\caption{\textbf{Parameters of the waveform relaxation algorithm.} The table
shows the different parameters of the waveform relaxation algorithm
together with their C++ data-type, default value, and a brief description.\label{tab:wfr-parameters}}
\end{table}

\prettyref{tab:wfr-parameters} summarizes the parameters of our reference
implementation of the waveform-relaxation technique. A subset was
previously introduced by \citet{Hahne15_00022} (\texttt{wfr\_interpolation\_order},
\texttt{wfr\_max\_iterations, wfr\_tol}), but we rename them here
to arrive at more descriptive names. The remaining parameters (\texttt{use\_wfr},
\texttt{wfr\_comm\_interval}) result from the generalization to rate-based
models.

\section{Results\label{sec:Results}}

In the following, we assess the stability of the different numerical
solution schemes and benchmark their performance on large-scale machines.
Furthermore, we illustrate the application of the simulation framework
to different models relevant in the neuroscientific literature.

\subsection{Stability and accuracy of integration methods\label{subsec:Comparison-of-numerical}}

To investigate the accuracy and stability of the different numerical
methods (see \prettyref{subsec:Approximate-numerical-solution}) we
consider an exactly solvable network of linear rate neurons with $\mu=0$
(see also \prettyref{subsec:Rate-models}) 

\begin{equation}
\tau dX(t)=A\cdot X(t)\,dt+\sqrt{\tau}\sigma\,dW(t)\,.\label{eq:lin-sde}
\end{equation}

\noindent The exact solution of the system of SDEs coincides with
the exponential Euler scheme and involves a matrix exponential and
a matrix square root \prettyref{eq:stochastic-exp-euler}. This exact
solution cannot be obtained with a distributed representation of $A$
as it is typically employed in the distributed simulation scheme of
a spiking network simulation code (see \prettyref{par:Restrictions}).
However, using the methods for numerical matrix computations in MATLAB
or Python \citep[both provide an implementation of the same state-of-the-art algorithms, see][]{Al-Mohy09,Deadman12},
we obtain an approximation to the exact solution, close to the general
limits of floating point numerics, and use this as a reference to
compute the root mean square error of the different approximative
methods. In order to employ the root mean square error in the context
of stochastic differential equations we compute the reference solution
for every tested step size and use the same random numbers for both
the reference solution and the approximative schemes. Furthermore,
we consider analytical stability criteria for some of the employed
methods. 

\noindent In the following we assume that $A$ is diagonalizable,
i.e. $A=T^{-1}DT$ with $T=(t^{ij})_{N\times N}\in\mathbb{C}^{N\times N}$
and $D=\mathrm{diag}(\lambda_{1},\ldots,\lambda_{N})$, and transform
the system of SDEs with $Z(t)=T\,X(t)$. It follows

\begin{eqnarray*}
\tau dZ(t) & = & D\cdot Z(t)\,dt+\sqrt{\tau}\sigma T\,dW(t)
\end{eqnarray*}

\noindent and $Z(t_{0})=TX_{0}$. The transformed system consists
of $N$ equations of the form

\begin{equation}
\tau dZ^{i}(t)=\lambda_{i}\cdot Z^{i}(t)\,dt+\sum_{j=1}^{N}\sqrt{\tau}\sigma t^{ij}\,dW^{j}(t)\quad\quad i=1,\ldots,n\label{eq:transformed-lin-sde}
\end{equation}

\noindent that depend on the eigenvalues of $A$ and are independent
of each other except for the contribution of the Wiener processes
$W^{j}(t)$ . For eigenvalues $\lambda_{i}\in\mathbb{C}$ with negative
real part $\mathrm{Re}(\lambda_{i})<0$, the solution of the $i$-th
transformed equation satisfies 

\[
\vert Z^{i}(t)-\tilde{Z}^{i}(t)\vert=e^{\lambda_{i}(t-t_{0})/\tau}\vert Z_{0}^{i}-\tilde{Z}_{0}^{i}\vert<\vert Z_{0}^{i}-\tilde{Z}_{0}^{i}\vert
\]

\noindent for two different initial values $Z_{0}^{i}$ and $\tilde{Z}_{0}^{i}$.
It is a desirable stability criterion that a numerical method applied
to \prettyref{eq:lin-sde} conserves this property. This is closely
related to the concept of A-stability for SDEs (see \citet{Kloeden92},
Chapter $9.8$) and A- respectively B-stability for ODEs \citep{Hairer91}.
A straight-forward calculation shows that the implicit Euler method
and the exponential Euler scheme retain this condition regardless
of the step size $\Delta t$ and that the Euler-Maruyama method retains
the condition if $\vert1+\lambda_{i}\cdot\Delta t/\tau\vert<1$ holds.
For $\lambda_{i}\in\mathbb{R}$ we obtain the step size restriction
$\Delta t<\frac{2\tau}{\vert\lambda_{i}\vert}$ and for complex eigenvalues
the condition is conserved if $\lambda_{i}\cdot\Delta t/\tau$ is
located inside a unit circle centered at $-1$ in the complex plane.

\begin{figure}[!tbh]
\begin{centering}
\includegraphics{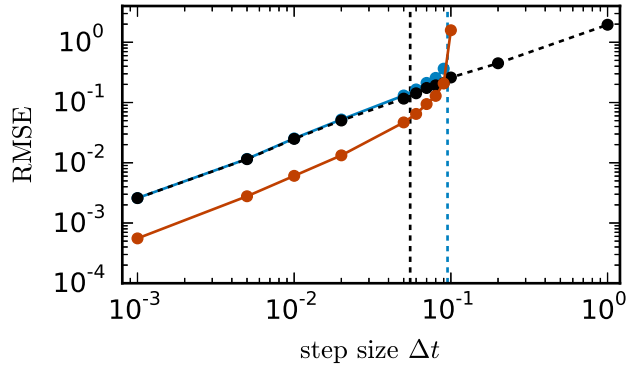}
\par\end{centering}
\caption{\textbf{Comparison of numerical methods for an all-to-all connected
inhibitory network.} $\mathrm{RMSE}=\sqrt{\frac{1}{N(t_{n}-t_{0})}\sum_{i=1}^{N}\sum_{j=1}^{n}(X_{j}^{i}-\widehat{X}_{j}^{i})^{2}}$
of the solution $X$ obtained by the approximate solvers (black dashed
curve: implicit Euler method solved with Newton iteration, blue curve:
Euler-Maruyama method, red curve: exponential Euler method) with respect
to the exact solution $\hat{X}$ as a function of step size in double
logarithmic representation. The black vertical line marks the largest
step size for which the implicit Euler method solved with fixed-point
iteration converges against the one obtained with Newton iteration.
The blue vertical line corresponds to the analytical stability restriction
$\Delta t\leq\frac{2}{21}$ of the Euler-Maruyama method. Network
parameters: size $N=400$, all-to-all connectivity with $w^{ij}=\frac{-1}{\sqrt{N}}$,
$\mu=0$, $\sigma=10$ and $\tau=1\protect\ms$. \label{fig:numerical-methods}}
\end{figure}

We investigate the accuracy of the numerical methods for an all-to-all
connected network with inhibitory connections of weight $w^{ij}=\frac{-1}{\sqrt{N}}$.
The eigenvalues of $A=-I+\frac{-1}{\sqrt{N}}\cdot\mathds{1}$ are
$\lambda_{1}=-1-\sqrt{N}$ and $\lambda_{2}=\ldots=\lambda_{N}=-1$.
It follows that the Euler-Maruyama scheme satisfies the stability
criterion for $\Delta t\leq\frac{2\tau}{\sqrt{N}+1}$. \prettyref{fig:numerical-methods}
shows the root mean square error of the different numerical schemes
for an example network. With decreasing step size all investigated
methods converge towards the exact solution with convergence order
$1$, which is consistent with the established theory for SDEs with
additive noise \citep{Kloeden92}. The Euler-Maruyama scheme only
works within the calculated stability region. The exponential Euler
with $A=-I$ and $f(X)=\frac{-1}{\sqrt{N}}\cdot\mathds{1}\cdot X$,
in the following called scalar exponential Euler, shows a similar
stability, as the whole input from the network is approximated with
an (explicit) Euler-like approach. Within the stability region of
Euler-Maruyama, however, the scalar exponential Euler yields more
accurate results than the two other methods. The implicit Euler scheme
solved with Newton iteration shows no stability issues, but it is
not applicable in the distributed simulation framework for spiking
neuronal networks (see \prettyref{par:Restrictions}). For completeness,
we also test the implicit Euler scheme with a parallelizable Jacobi
fixed-point iteration. The convergence properties of the fixed-point
iteration demand the scheme to be contractive \citep[see e.g.][ Sec. 4.2]{Kelley95}.
Therefore, in our test case the step size is restricted to roughly
$\Delta t<0.05\ms$ and accordingly to a region where the scalar exponential
Euler yields better results. In addition, an iterative scheme in each
single time step is expected to be much more time consuming than using
the scalar exponential Euler scheme. Based on these results we employ
the scalar exponential Euler to solve rate-based neuron dynamics \prettyref{eq:network-OUP},
as it is the most accurate, stable and efficient scheme compatible
with the constraints of the distributed simulation scheme for spiking
neural networks. Nevertheless, the inevitable restrictions on the
step size $\Delta t$ need to be taken into account in order to obtain
an accurate solution. An appropriate step size can be estimated with
the analytical stability criterion of the Euler-Maruyama method. For
an all-to-all connected inhibitory network the restriction $\Delta t\leq\frac{2\tau}{\sqrt{N}+1}$
shows that, with increasing network size $N$ or decreasing time constant
$\tau$ the step size $\Delta t$ needs to be reduced.

\begin{figure}[!tbh]
\begin{centering}
\includegraphics{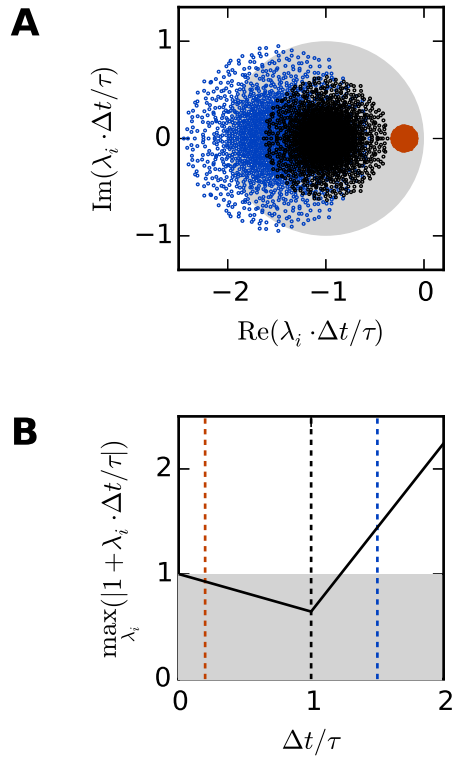}
\par\end{centering}
\caption{\textbf{Stability analysis for a balanced excitatory-inhibitory network.}
The network contains in total $N$ neurons where the number of excitatory
neurons is four times larger then the number of inhibitory neurons.
Each neuron receives input from a fixed number of $0.8\cdot p\cdot N$
excitatory and $0.2\cdot p\cdot N$ inhibitory randomly chosen source
neurons with connection probability $p$ and connection weights $\frac{1}{\sqrt{N}}$
and $\frac{-4}{\sqrt{N}}$, respectively\textbf{. (A)} Black circles
show the eigenvalues $\lambda_{i}$ of the matrix $A$ for a network
of $N=2000$ neurons. Blue and red circles show the rescaled eigenvalues
$\lambda_{i}\cdot\Delta t/\tau$ for $\Delta t/\tau=1.5$ and $\Delta t/\tau=0.2$.
The filled gray circle indicates the region where the rescaled eigenvalues
$\lambda_{i}\cdot\Delta t/\tau$ meet the stability criterion $\vert1+\lambda_{i}\cdot\Delta t/\tau\vert<1$
of the Euler-Maruyama method. \textbf{(B)} The curve shows the maximum
of $\vert1+\lambda_{i}\cdot\Delta t/\tau\vert$ over all eigenvalues
$\lambda_{i}$ dependent on $\Delta t/\tau$. The gray area again
indicates the region where the stability criterion of the Euler Maruyama
method is meet. Colored vertical lines correspond to the rescaled
eigenvalues displayed in panel (A). \label{fig:eigenvalues-of-network}}
\end{figure}

The fully connected network constitutes the worst case test for the
class of rate-based models \prettyref{eq:network-OUP}, as the absolute
value of the negative eigenvalue quickly increases with the number
of neurons $N$. A network which does not suffer from this problem
is a perfectly balanced network of excitatory and inhibitory neurons
\citep{Rajan06}. For these models the conditions on the step size
$\Delta t$ of the Euler-Maruyama method are less restrictive and
the same is expected for the scalar exponential Euler method. As an
example we employ a sparse balanced excitatory-inhibitory-network.
In a scaling of the connection weights as $\frac{1}{\sqrt{N}}$, the
spectral radius of $A$ and therefore the subsequent stability analysis
is independent of $N$. \prettyref{fig:eigenvalues-of-network}B demonstrates
that for this test case the Euler-Maruyama method is stable for $\Delta t<1.2\tau$.
Given a commonly used simulation step size of $h=\Delta t=0.1$, networks
of this kind can be safely simulated if the time constant $\tau$
fulfills $\tau\geq0.085$.

Random networks with incomplete balance exhibit both types of stability
issues discussed above. In this case the matrix $A$ contains an eigenvalue
$\lambda_{1}=-1-\rho\sqrt{N}$ which scales with the network size,
however, with a proportionality constant $|\rho|<1$ which is reduced
compared to the fully connected inhibitory network and determined
by the sparseness and the partial balance. Nevertheless, the network
size needs to be taken into account for the choice of the step size.
The latter also needs to ensure that the cloud of eigenvalues determined
by the randomness in the connectivity meets the stability criterion.

\subsection{Performance of iteration schemes\label{subsec:Performance}}

This section investigates the performance of the rate model implementation.
We are interested in i) the scalability of the rate model framework
and ii) the comparison between the straight-forward implementation
with communication in every computation time step and the iterative
approach using waveform relaxation (see \prettyref{par:Waveform-relaxation-techniques}
for details). We perform the simulations on the JUQUEEN BlueGene/Q
supercomputer \citep{stephan2015juqueen} at the Jülich Research Centre
in Germany. It comprises $28,672$ compute nodes, each with a $16$-core
IBM PowerPC A2 processor running at 1.6 GHz. For our benchmarks we
use 8 OpenMP threads per JUQUEEN compute node and denote by $\mathrm{VP}=8\cdot\#\mathrm{nodes}$
the total number of virtual processes employed. 

As a test case we employ the excitatory-inhibitory-network of linear
rate neurons ($\phi(x)=\psi(x)=x$) introduced in \prettyref{subsec:Comparison-of-numerical},
but with a fixed number of inputs ($2000$) independent of the number
of neurons to allow for an unbiased weak scaling. 

\begin{figure}[!tbh]
\begin{centering}
\includegraphics{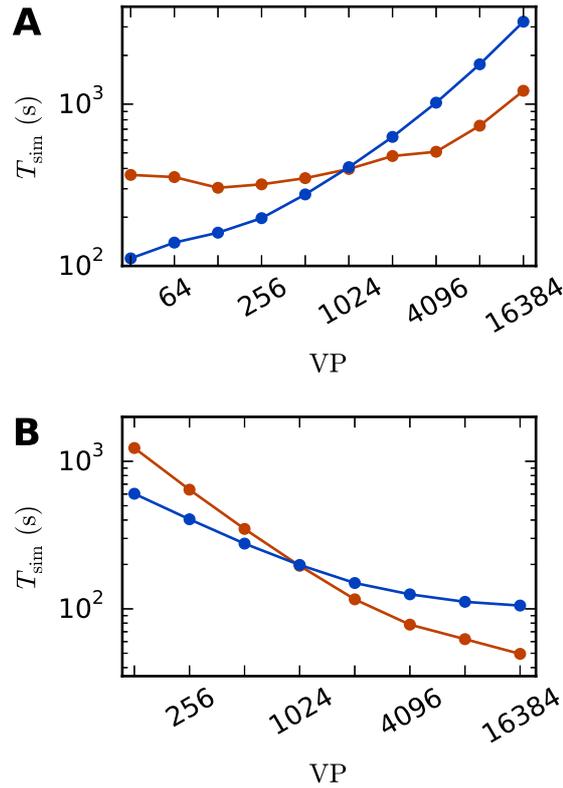}
\par\end{centering}
\caption{\textbf{Scaling behavior of an excitatory-inhibitory network.} Simulation
time with waveform relaxation (red curves\texttt{, wfr\_comm\_interval}:
$1.0\protect\ms$, \texttt{wfr\_tol}: $10^{-4}$) and without waveform
relaxation (blue curves) as a function of the number of virtual processes
in double logarithmic representation. The simulations span $T=100\protect\ms$
of biological time at a computation step size of $h=0.1\,\text{ms}$.\textbf{
(A)} Weak scaling with $100$ neurons per virtual process $\mathrm{VP}$.
\textbf{(B)} Strong scaling with a total number of $N=51,200$ neurons.
Other network parameters as in \prettyref{subsec:Comparison-of-numerical}
with $\mu=0$, $\sigma=1$ and $\tau=10\,\text{ms}$.\label{fig:Network-behaviour}}
\end{figure}

A weak scaling (\prettyref{fig:Network-behaviour}A) shows that the
scalability of the straight-forward computation is impaired by the
massive amount of communication. While for perfect scaling the simulation
time should be constant over the number of virtual processes, the
actual simulation time is increased by $15-25\%$ when the number
of virtual processes is doubled for $\mathrm{VP}<256$ and even up
to $83\%$ from $8,192$ to $16,384$ virtual processes. For the iterative
method, the scaling behavior is close to constant up to $1,024$ virtual
processes. When more processes are employed, the simulation time is
increasing. However, the iterative method shows a better scaling behavior
as the increase is weaker compared to the straight-forward computation
due to the lower total number of communication steps. Due to the higher
computational load of the iterative method (see \prettyref{par:Waveform-relaxation-techniques})
the simulation time is larger compared to the straight forward approach
for a small number of $\mathrm{VP}$, where communication is not that
crucial. For $\mathrm{VP}\ge1024$, the iterative approach is superior
with a speed up factor close to three for $16,384$ virtual processes
($1209\,\mathrm{s}$ vs. $3231\,\mathrm{s}$).

The strong scaling scenario with a fixed total number of $N=51,200$
neurons in \prettyref{fig:Network-behaviour}B constitutes a similar
result. The iterative approach is beneficial for more than $1,024$
virtual processes and the scaling behavior of the iterative method
outperforms that of the straight-forward computation. Starting at
$4,096$ virtual processes the savings in computation time decrease,
which is explained by the very low workload of each single compute
node. Again, for a smaller number of virtual processes the amount
of additional computations is too high to outperform the straight-forward
computation.

Despite the overall good scaling behavior, the performance in terms
of absolute compute time is inferior to a simulator specifically designed
for rate-based models alone (not shown). In the latter case it increases
performance to collect the states of all neurons in one vector.
If further the connectivity is available in form of a matrix and the
delays are zero or homogeneous, the network can be efficiently updated
with a single matrix-vector multiplication. Thus the increased functionality
and flexibility of having rate- and spiking model neurons unified
in one simulator comes for the price of a loss of performance for
the rate-based models. However, as noted in the introduction, the
number of neurons in rate-based network models is usually small and
therefore performance is not as critical as for spiking network models.

\subsection{Applications\label{subsec:Applications}}

This section presents several applications of the simulation framework.
First, we discuss a balanced random network of linear rate units,
then include nonlinear neuron dynamics in a random network, and spatially
structured connectivity in a functional neural-field model. In each
case, simulation results are compared to analytical predictions. Furthermore,
we simulate a mean-field model of a spiking model of a cortical microcircuit
and discuss possible generalizations.

\subsubsection{Linear model\label{subsec:Linear-model}}

In the asynchronous irregular regime which resembles cortical activity,
the dominant contribution to correlations in networks of nonlinear
units is given by effective interactions between linear response modes
\citep{Grytskyy13_131,Trousdale12_e1002408,Pernice11_e1002059,Dahmen16_031024}.
Networks of such noisy linear rate models have been investigated to
explain features such as oscillations \citep{Bos2015_arxiv} or the
smallness of average correlations \citep{Tetzlaff12_e1002596,Helias13_023002}.
We here consider a prototypical network model of excitatory and inhibitory
neurons following the linear dynamics given by \eqref{eq:network-OUP}
with $\phi(x)=\psi(x)=x$, $\mu=0$, and noise amplitude $\sigma$,
\begin{align}
\tau dX^{i}(t) & =\left(-X^{i}+\sum_{j=1}^{N}w^{ij}X^{j}(t)\right)dt+\sqrt{\tau}\sigma dW^{i}(t)\,.
\end{align}
Due to the linearity of the model, the cross-covariance between neurons
$i$ and $j$ can be calculated analytically and is given by \citep{Ginzburg94,Risken96,Gardiner04,Dahmen16_031024}
\begin{equation}
c(t)=\sum_{i,j}\frac{v^{i\mathrm{T}}\sigma^{2}v^{j}}{\lambda_{i}+\lambda_{j}}u^{i}u^{j\mathrm{T}}\left(\theta(\Delta)\frac{1}{\tau}e^{-\lambda_{i}\frac{t}{\tau}}+\theta(-\Delta)\frac{1}{\tau}e^{\lambda_{j}\frac{t}{\tau}}\right),
\end{equation}
where $\theta$ denotes the Heaviside function. The $\lambda_{i}$
indicate the eigenvalues of the matrix $1-W$ corresponding to the
$i$-th left and right eigenvectors $v^{i}$ and $u^{i}$ respectively.
Non-zero delays yield more complex analytical expressions for cross-correlations.
In the population-averaged case, theoretical predictions are still
analytically tractable \citep[eq. 18 in][]{Grytskyy13_131}. 
\begin{figure}[t]
\begin{centering}
\includegraphics{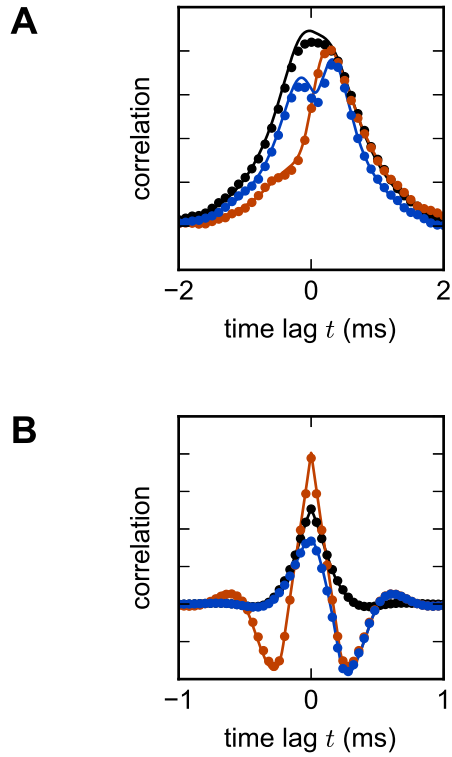}
\par\end{centering}
\caption{\textbf{Linear rate model of a random excitatory-inhibitory network.}
\textbf{(A)} Cross-correlation functions of two pairs of excitatory
neurons (black, red) and an excitatory-inhibitory neuron pair (blue)
in a network without delay. The variability across correlation functions
arises from heterogeneity in network connections (difference between
black and red curves) and from different combinations of cell types
(e.g. difference between black and blue curves). \textbf{(B)} Population-averaged
autocorrelation function for excitatory (black) and inhibitory neurons
(red), and cross-correlation function between excitatory and inhibitory
neurons (blue) in a network with delay $d=2\protect\ms$. Symbols
denote simulation results, curves show theoretical predictions. Parameters:
$N_{E}=80$ excitatory and $N_{I}=20$ inhibitory neurons, random
connections with fixed out-degree, connection probability $p=0.1$,
excitatory weight $w_{E}=1/\sqrt{N_{E}+N_{I}}$, inhibitory weight
$w_{I}=-6\,w_{E}$, $\tau=1\protect\ms$, $\mu=0$, $\sigma=1.$ \label{fig:Lin-rate-model}}
\end{figure}
\prettyref{fig:Lin-rate-model} shows the cross-covariance functions
for pairs of instantaneously coupled neurons in a large network, as
well as population-averaged covariance functions in a network of excitatory
and inhibitory neurons with delayed interactions. In both cases, simulations
are in good agreement with the theoretical predictions. 

\subsubsection{Nonlinear model}

\begin{figure}[t]
\centering{}\includegraphics{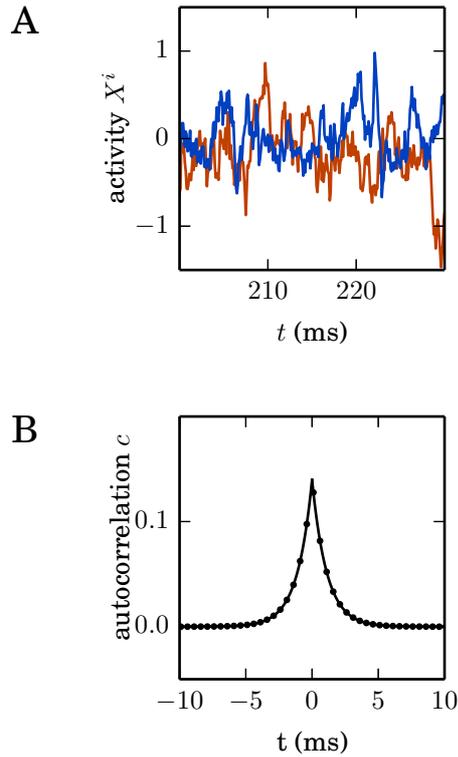}\caption{\label{fig:scs}\textbf{Nonlinear network model. }Simulation of the
network specified by \ref{eq:diffeq_motion} with $N=1000$ neurons.
\textbf{(A) }Noisy example trajectories of two neurons. \textbf{(B)
}Autocorrelation function obtained by simulation averaged over all
neurons (dots) and theory (solid curve). Other parameters: $\tau=1\protect\ms$,
$\sigma=0.5$, $g=0.5$. Step size as in \prettyref{fig:Network-behaviour}.}
\end{figure}

So far we considered a network with linear couplings between the units.
Qualitatively new features appear in the presence of nonlinearities.
One of the most prominent examples is the emergence of chaotic dynamics
\citep{Sompolinsky88_259} in a network of non-linearly coupled rate
units. The original model is deterministic and has been recently extended
to stochastic dynamics \citep{Goedeke16_arxiv}. The model definition
follows from \prettyref{eq:network-OUP} with $\mu=0,\,\phi(x)=\mathrm{x},\,\psi(x)=\tanh(x)$,
i.e.

\begin{align}
\tau dX^{i}(t) & =\left(-X^{i}(t)+\sum_{j=1}^{N}w^{ij}\tanh\left(X^{j}(t)\right)\right)dt+\sqrt{\tau}\sigma\,dW^{i}(t)\,,\label{eq:diffeq_motion}
\end{align}
where $w^{ij}\approx\mathcal{N}(0,g^{2}/N)$ are Gaussian random couplings.
In the thermodynamic limit $N\rightarrow\infty$, the population averaged
autocorrelation function $c(t)$ can be determined within dynamic
mean-field theory \citep{Sompolinsky88_259,Goedeke16_arxiv,Schuecker16b_arxiv}.
Comparing $c(t)$ obtained by simulation of a network \prettyref{eq:diffeq_motion}
with the analytical result \citep[eqs. 6 and 8]{Goedeke16_arxiv}
demonstrates excellent agreement (\prettyref{fig:scs}). The simulated
network is two orders of magnitude smaller than the cortical microcircuit,
illustrating that in this context finite-size effects are already
negligible at this scale.

\subsubsection{Functional model}

Complex dynamics not only arises from nonlinear single-neuron dynamics,
but also from structured network connectivity \citep{Yger2011_229}.
One important nonrandom feature of brain connectivity is the spatial
organization of connections \citep{Malach93,Voges10_277}. In spatially
structured networks, delays play an essential role in shaping the
collective dynamics \citep{Roxin05,Voges12}. Patterns of activity
in such networks are routinely investigated using neural-field models.
In contrast to the models discussed above, field models require a
discretization of space for numerical simulation. Such discretization
can be done in the real space, leading effectively to a network of
neurons at discrete positions in space, or alternatively, for particular
symmetries in the couplings, in k-space \citep{Roxin06}. Here, we
follow the more general approach of discretization in real space.

A prototypical model of a spatial network is given by \citep{Roxin05},
where the authors consider the neural-field model
\begin{equation}
\tau dX(\varphi,t)=\left(-X(\varphi,t)+\phi\left[I_{\mathrm{ext}}+\int_{-\pi}^{\pi}d\varphi^{\prime}\,w(|\varphi-\varphi^{\prime}|)X(\varphi^{\prime},t-d)\right]\right)\,dt\label{eq:neural-field}
\end{equation}
with delayed (delay $d$) interactions, constant input $I_{\mathrm{ext}}$,
threshold-linear activation function $\phi=x\cdot\theta(x)$ and periodic
Mexican-hat shaped connectivity 
\begin{equation}
w(|\varphi-\varphi^{\prime}|)=w_{0}+w_{1}\cos(\varphi-\varphi^{\prime}).
\end{equation}
The spatial variable $\varphi$ can also be interpreted as the preferred
orientation of a set of neurons, thus rendering \prettyref{eq:neural-field}
a model in feature space \citep{Hansel98a}. Discretizing space into
$N$ segments yields the following set of coupled ODEs
\begin{equation}
\tau dX^{i}=\left(-X^{i}+\phi\left[I_{\mathrm{ext}}+\sum_{j=1}^{N}w^{ij}X^{j}(t-d)\right]\right)dt
\end{equation}
with connectivity $w^{ij}=\frac{2\pi}{N}w(|\varphi^{i}-\varphi^{j}|)$,
$\varphi_{i}=-\pi+\frac{2\pi}{N}\cdot i$ for $i\in[1,N]$ and discretization
factor $\frac{2\pi}{N}$ that scales the space constants $w_{0}$
and $w_{1}$ with the neuron density. The spatial connectivity together
with a delay in the interaction introduce various spatial activity
patterns depending on the shape of the Mexican-hat connectivity. 
\begin{figure}[t]
\begin{centering}
\includegraphics{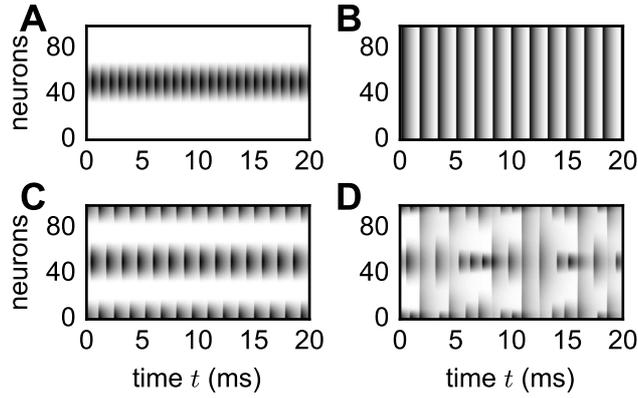}
\par\end{centering}
\caption{\textbf{Spatial patterns in functional neural-field model.} Vertical
axis shows neuron indices organized according to ascending angle $\varphi\in[-\pi,\pi)$.
Neuronal activity $X_{i}(t)=X(\varphi_{i},t)$ encoded by gray scale
with white denoting no activity. Initial transients not shown. Patterns
reproduce the analytically derived phase diagram in the original study
by \citet{Roxin05}. Parameters: $N=100$, $d=0.1\protect\ms$, $\tau=1\protect\ms$,
$I_{\mathrm{ext}}=1$, $w_{0}=-80$, $w_{1}=15$ (\textbf{A}), $w_{1}=5$
(\textbf{B}), $w_{1}=-46$ (\textbf{C}), $w_{1}=-86$ (\textbf{D}).
Initial condition: $X_{i}(0)=X(\varphi_{i},0)=\pi^{2}-\varphi_{i}^{2}$.
\label{fig:spatial_net}}
\end{figure}
To illustrate applicability of the simulation framework to neural-field
models, we reproduce various patterns (\prettyref{fig:spatial_net})
observed by \citet{Roxin05}. Although the discrete and continuous
networks strictly coincide only in the thermodynamic limit $N\rightarrow\infty$,
numerically obtained patterns shown in \prettyref{fig:spatial_net}
well agree with the analytically derived phase diagram of the continuous
model \citep{Roxin05} already for network sizes of only $N=100$
neurons. 

\subsubsection{Mean-field analysis of complex networks\label{subsec:Mean-field-analysis-of}}

A network of spiking neurons constitutes a high dimensional and complex
system. To investigate its stationary state, one can describe the
activity in terms of averages across neurons and time, leading to
population averaged stationary firing rates \citep{Brunel00}. Here,
the spatial average collapses a large number of neurons into a single
population, which interpreted as a single rate unit. The ability to
represent spiking as well as rate dynamics by the same simulation
framework allows a straight-forward analysis of the spiking network
by replacing the spiking neuron populations by single rate-based model
neurons.

In more formal terms, we now consider networks of neurons structured
into $N$ interconnected populations. A neuron in population $\alpha$
receives $K_{\alpha\beta}$ incoming connections from neurons in population
$\beta$, each with synaptic efficacy $w^{\alpha\beta}$. Additionally,
each neuron in population $\alpha$ is driven by $K_{\alpha\mathrm{,ext}}$
Poisson sources with rate $X_{\mathrm{ext}}$ and synaptic efficacy
$w_{\mathrm{ext}}$. We assume leaky integrate-and-fire model neurons
with exponentially decaying post-synaptic currents. The dynamics of
membrane potential $V$ and synaptic current $I_{\mathrm{s}}$ is
\citep{Fourcaud02}

\begin{eqnarray}
\tau_{m}\frac{dV^{i}}{dt} & = & -V^{i}+I_{s}^{i}\nonumber \\
\tau_{s}\frac{dI_{s}^{i}}{dt} & = & -I_{s}^{i}+\tau_{m}\sum_{j=1}^{N}w^{ij}\sum_{k}\delta(t-t_{k}^{j}-d)\,,\label{eq:Inestrescaled}
\end{eqnarray}
where $t_{k}^{j}$ denotes the $k$-th spike-time of neuron $j$,
and $\tauM$ and $\tau_{\mathrm{s}}$ are the time constants of membrane
and synapse, respectively. The membrane resistance has been absorbed
in the definition of the current. Whenever the membrane potential
$V$ crosses the threshold $\theta,$ the neuron emits a spike and
$V$ is reset to the potential $V_{\mathrm{r}}$, where it is clamped
for a period of length $\taur$. If we assume that all neurons have
identical parameters, a diffusion approximation leads to the population-averaged
firing rates $X_{\alpha}$ \citep{Fourcaud02} \foreignlanguage{english}{:=}
\begin{eqnarray}
\frac{1}{X_{\alpha}} & = & \taur+\taum\sqrt{\pi}\int_{(V_{\mathrm{r}}-\mu_{\alpha})/\sigma_{\alpha}+\gamma\sqrt{\taus/\taum}}^{(\theta-\mu_{\alpha})/\sigma_{\alpha}+\gamma\sqrt{\taus/\taum}}\,e^{u^{2}}\ (1+\erf(u))\,\d u\nonumber \\
 & =: & 1/\Phi_{\alpha}(X)\label{eq:siegert_2D}\\
\mu_{\alpha} & = & \taum\sum_{\beta}K_{\alpha\beta}w_{\alpha\beta}X_{\beta}+\taum K_{\mathrm{\alpha,ext}}w_{\mathrm{ext}}X_{\mathrm{ext}}\label{eq:mean}\\
\sigma_{\alpha}^{2} & = & \taum\sum_{\beta}K_{\alpha\beta}w_{\alpha\beta}^{2}X_{\beta}+\taum K_{\mathrm{\alpha,ext}}w_{\mathrm{ext}}^{2}X_{\mathrm{ext}}.\label{eq:sigma}
\end{eqnarray}
Here, $\gamma=|\zeta(1/2)|/\sqrt{2},$ with $\zeta$ denoting the
Riemann zeta function \citep{Abramowitz74}. We find the fixed points
of \prettyref{eq:siegert_2D} by solving the first-order differential
equation \citep{Wong_06,Schuecker16_arxiv_v3}

\begin{equation}
\tau\d X^{\alpha}=\left(-X^{\alpha}+\Phi(X)\right)dt,\label{eq:int_siegert}
\end{equation}
which constitutes a network of rate neurons with the dimension equal
to the number of populations $N$. 

\begin{figure}[t]
\begin{centering}
\includegraphics{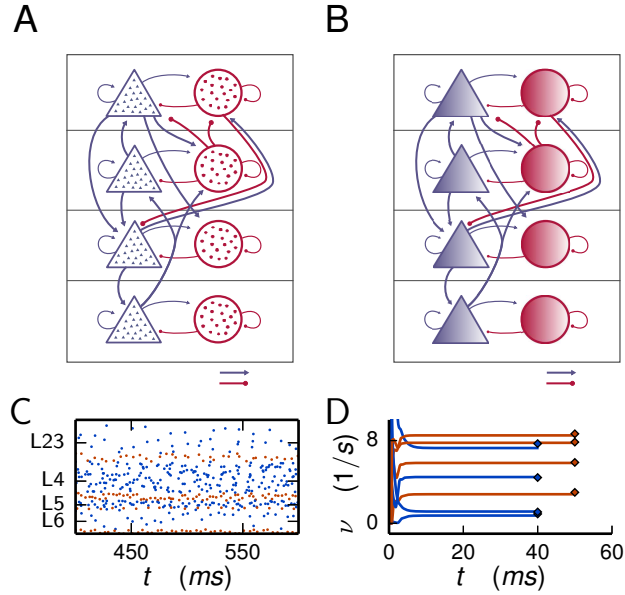}
\par\end{centering}
\caption{\label{fig:microcircuit}\textbf{Reduction of spiking microcircuit
model to rate dynamics}. \textbf{(A)} Sketch of a microcircuit model
\citep{Potjans14_785} with excitatory (blue triangles) and inhibitory
(red circles) neuron populations, each consisting of a large number
of neurons indicated by the small triangles and disks respectively.
Arrows between populations indicate the in-degree \textbf{$K$}.\textbf{
(B)} Sketch of the corresponding reduced model where each population
is replaced by a single rate unit. \textbf{(C)} Spiking activity in
the different layers. \textbf{(D)} Dynamics of the eight neurons of
the rate network \prettyref{eq:int_siegert} (curves) and comparison
to population averaged firing rates obtained from direct simulations
of the spiking network (diamonds).}
\end{figure}

Next we apply this framework to a cortical microcircuit model \citep{Potjans14_785}
constituting roughly $80000$ spiking neurons structured into $8$
populations across $4$ layers $[L23,L4,L5,L6]$, with one excitatory
and one inhibitory cell type each (\prettyref{fig:microcircuit}).
The model exhibits irregular and stationary spiking activity (\prettyref{fig:microcircuit}C).
Replacing each population by a single rate neuron (\ref{fig:microcircuit}B)
results in an eight-dimensional rate network \prettyref{eq:int_siegert}
which converges to a fixed point corresponding to the population-averaged
firing rates obtained from direct simulation of the spiking model
(\prettyref{fig:microcircuit}D).

\subsubsection{Further neuron models\label{subsec:Other-neuron-models}}

\paragraph{Nonlinear neuron dynamics}

A common feature of various rate neurons considered so far is the
leaky neuron dynamics, i.e the linear term $-X^{i}(t)$ in \prettyref{eq:network-OUP}.
However the presented framework can be easily extended to nonlinear
neuron dynamics as used for example by \citet{Stern14_062710}. In
a more general form \prettyref{eq:network-OUP} reads

\begin{equation}
\tau dX^{i}(t)=\left[a(X^{i}(t))+\phi\left(\sum_{j=1}^{N}w^{ij}\psi\left(X^{j}(t-d)\right)\right)\right]\,dt+\sqrt{\tau}\sigma dW^{i}(t)\label{eq:network-nonlinear}
\end{equation}
where $a$ characterizes the intrinsic neuron dynamics. If $a$ does
not contain a linear part, the Euler-Maruyama scheme can be used for
the update, i.e,

\begin{equation}
X_{k+1}^{i}=X_{k}^{i}+\left[a(X_{k}^{i})+\mu+\phi\left(\sum_{j=1}^{N}w^{ij}\psi\left(X_{k-\frac{d}{\Delta t}}^{j}\right)\right)\right]\,\frac{1}{\tau}\Delta t+\frac{1}{\sqrt{\tau}}\sigma\Delta W_{k}^{i}.\label{eq:euler_maruyama_nonlinear}
\end{equation}
If $a$ also contains a linear part, so that $a(X^{i})=-X^{i}+f(X^{i})$,
one can use an exponential Euler update approximating the non-linear
part as constant during the update. This leads to

\begin{equation}
X_{k+1}^{i}=e^{-\Delta t/\tau}X_{k}^{i}+\left(1-e^{-\Delta t/\tau}\right)\left[f(X_{k}^{i})+\phi\left(\sum_{j=1}^{N}w^{ij}\psi\left(X_{k-\frac{d}{\Delta t}}^{j}\right)\right)\right]+\sqrt{\dfrac{1}{2}(1-e^{-2\Delta t/\tau})}\,\sigma\eta_{k}^{i},\label{eq:exponential_euler_nonlinear}
\end{equation}
with $\eta_{k}^{i}\sim\mathcal{N}(0,1).$

\paragraph{Multiplicative coupling}

Another possible extension is a multiplicative coupling between units
as for example employed in \citet{Gancarz98_1159} or the original
works of \citet{Wilson72_1,Wilson73}. In the most general form, this
amounts to

\begin{equation}
\tau dX^{i}(t)=\left[-X^{i}(t)+H(X^{i})\cdot\phi\left(\sum_{j=1}^{N}w^{ij}\psi\left(X^{j}(t-d)\right)\right)\right]\,dt+\sqrt{\tau}\sigma dW^{i}(t)\,,\label{eq:network-multiplicative-coupling}
\end{equation}
which, again assuming the coupling term to be constant during the
update, can be solved using the exponential Euler update

\begin{equation}
X_{k+1}^{i}=e^{-\Delta t/\tau}X_{k}^{i}+\left(1-e^{-\Delta t/\tau}\right)\left[H(X_{k}^{i})\cdot\phi\left(\sum_{j=1}^{N}w^{ij}\psi\left(X_{k-\frac{d}{\Delta t}}^{j}\right)\right)\,\right]+\sqrt{\dfrac{1}{2}(1-e^{-2\Delta t/\tau})}\,\sigma\eta_{k}^{i},\label{eq:exponential_euler_multiplicative}
\end{equation}
with $\eta_{k}^{i}\sim\mathcal{N}(0,1).$

\paragraph{Multiplicative noise}

So far, we have considered rate models subject to additive noise corresponding
to $b(t,x(t))=b(t)$ in \prettyref{eq:system-sdes}. The linear rate
model considered in \prettyref{subsec:Linear-model} describes the
dynamics around a stationary state and due to the stationary baseline,
the noise amplitude is constant. However, one might relax the stationarity
assumption which would render the noise amplitude proportional to
the time dependent rate, i.e. a multiplicative noise amplitude. The
presented framework covers the latter since the exponential Euler
update is also valid for multiplicative noise \prettyref{eq:stochastic-exp-euler}.

\paragraph{Output noise}

\citet{Grytskyy13_131} show that there is a mapping between a network
of leaky integrate-and-fire models and a network of linear rate models
with so-called output noise. Here the noise is added to the output
rate of the afferent neurons 
\begin{equation}
\tau\frac{dX^{i}(t)}{dt}=-X^{i}(t)+\mu+\phi\left(\sum_{j=1}^{N}w^{ij}\psi\left(X^{j}(t-d)+\sqrt{\tau}\sigma\xi^{j}(t)\right)\right)\quad\quad i=1,\ldots,N\label{network-opn}
\end{equation}
and we cannot write the system as a SDE of type \prettyref{eq:general-sde},
as the nonlinearities $\phi(x)$ and $\psi(x)$ are also applied to
the white noise $\xi^{j}$. In addition to the implementation \texttt{rate\_neuron\_ipn}
for the rate-based models \prettyref{eq:network-OUP} discussed in
the present work, our reference implementation also contains a base
implementation \texttt{rate\_neuron\_opn} for models with output noise.
For these models, the stochastic exponential Euler method can not
be employed. Instead the solver assumes the noise $\xi^{j}$ to be
constant over the update interval which leads to the update formula
\begin{equation}
X_{k+1}^{i}=e^{-\Delta t/\tau}X_{k}^{i}+\left(1-e^{-\Delta t/\tau}\right)\left[\mu+\phi\left(\sum_{j=1}^{N}w^{ij}\psi\left(X_{k}^{j}+\sqrt{\frac{\tau}{\Delta t}}\sigma\eta_{k}^{j}\right)\right)\right]\,.\label{eq:update-opn}
\end{equation}

\noindent The term $X_{k}^{j}+\sqrt{\frac{\tau}{\Delta t}}\sigma\eta_{k}^{j}$
with $\eta_{k}^{j}\sim\mathcal{N}(0,1)$ is calculated beforehand
in the sending neuron $j$, which results in the same amount of communicated
data as in the case of models with input noise.

\section{Discussion\label{sec:Discussion}}

This work presents an efficient way to integrate rate-based models
in a neuronal network simulator that is originally designed for models
with delayed spike-based interactions. The advantage of the latter
is a decoupling of neuron dynamics between spike events. This is used
by current parallel simulators for large-scale networks of spiking
neurons to reduce communications between simulation processes and
significantly increase performance and scaling capabilities up to
supercomputers \citep{Morrison05a}. In contrast, rate-based models
interact in a continuous way. For delayed interactions, neuronal dynamics
are still decoupled for the minimal delay of the network such that
information can be exchanged on a coarse time-grid. For instantaneous
coupling, communication in every time step is required. This is feasible
for small networks that can be simulated on small machines and thus
require only a small amount of communication. For improved efficiency
of simulations of large networks on supercomputers, we implement an
iterative numerical solution scheme \citep{Lelarasmee82}. Furthermore,
we investigate several standard methods for the solution of rate model
equations and demonstrate that the scalar exponential Euler method
is the best choice in the context of a neuronal network simulator
that is originally designed for models with delayed spike-based interactions.
Afterwards, we show the applicability of the numerical implementation
to a variety of well-known and widely-used neuron models (\prettyref{subsec:Applications})
and illustrate possible generalizations to other categories of rate-based
neuron models.

The current reference implementation uses an exponential Euler scheme
\citep{Adamu11,Komori14} with a diagonal matrix $A$: The additive
noise as well as the leaky dynamics of single neurons are exactly
integrated while the input to the neurons is approximated as piecewise
constant. The analysis in \prettyref{subsec:Comparison-of-numerical}
demonstrates that the scalar exponential Euler is the most accurate
and stable standard-method for SDEs that is applicable to a distributed
spiking simulator. In particular the distributed design renders implicit
methods less feasible, as the convergence of the involved fixed-point
iteration requires small time-steps in a region where the exponential
Euler is already stable and more accurate. As the computation step
size needs to be compared against the time constant $\tau$, stable
solutions for small values $\tau\ll1$ may require to decrease the
step size below a default value. 

The reference implementation provides an optional iterative method,
the waveform relaxation \citep{Lelarasmee82}, for networks with instantaneous
rate connections. This method improves scalability by reducing communication
at the cost of additional computations. As a consequence, the optimal
method (standard vs. iterative) depends on the numbers of compute
nodes and virtual processes. In our test case the use of the waveform-relaxation
technique is beneficial for $1024$ or more virtual processes. It
is therefore recommended to employ the iterative scheme for large-scale
simulations on supercomputers, but to disable it for smaller rate-model
simulations on local workstations or laptops. This can easily be expressed
by the parameter \texttt{use\_wfr} (see \prettyref{par:Waveform-relaxation-techniques}
for details) of the algorithm. In general, the scalability for simulations
of rate models is worse than for spiking network simulations \citep{Kunkel14_78}
and comparable to simulations with gap junctions \citep{Hahne15_00022}.
This is expected since for rate neurons as well as for gap junctions
a large amount of data needs to be communicated compared to a spiking
simulation. Future work should assess whether this bottleneck can
be overcome by a further optimized communication scheme.

While our reference implementation uses the simulation software NEST
as a platform, the employed algorithms can be ported to other parallel
spiking network simulators. Furthermore, the implementation of the
example neuron models as C++ templates allows easy customization to
arbitrary gain functions. Researchers can easily create additional
models, without the need for in-depth knowledge of simulator specific
data structures or numerical methods. In addition, the developed infrastructure
is sufficiently general to allow for extensions to other categories
of neuron models as shown explicitly for nonlinear neuron dynamics,
multiplicative coupling, and other types of noise. This design enables
the usage of the framework for a large body of rate-based neural-network
models. Furthermore, the generality of the model equations allows
applications beyond neuronal networks, such as e.g. in computational
gliascience \citep{Amiri12_60} or artificial intelligence \citep{Haykin08}. 

Mean-field theory has built a bridge between networks of spiking neurons
and rate-based units that either represent single neurons or populations
\citep{Buice07_031118,Buice09_377,Ostojic11_e1001056,Bressloff12,Grytskyy13_131}.
In the latter case, the rate-based approach comes along with a considerable
reduction of the dimensionality (\prettyref{subsec:Mean-field-analysis-of}).
Due to a possibly large number of populations, the fixed-point solution
of the stationary activity can generally not be determined analytically,
but it can be found by evolving a pseudo-time dynamics. Within the
presented framework, this approach is much faster than the spiking
counter-part and thus can facilitate calibration of large-scale spiking
network models \citep{Schuecker16_arxiv_v3}.

The presented unifying framework allows researchers to easily switch
between rate-based and spiking neurons in a particular network model
requiring only minimal changes to the simulation script. This facilitates
an evaluation of the different model types against each other and
increases reproducibility in the validation of mean-field reductions
of spiking networks to rate-based models. Furthermore, it is instructive
to study whether and how the network dynamics changes with the neuron
model \citep{Brette15_151}. In particular, functional networks being
able to perform a given task are typically designed with rate-based
neurons. Their validity can now be evaluated by going from a more
abstract rate-based model to a biologically more realistic spiking
neuron model. The present reference implementation does not allow
for interactions between spiking and rate-based units. While this
is technically trivial to implement, the proper conversion from spikes
to rates and vice versa is a conceptual issue that has to be explored
further by theoretical neuroscience.

The presented joined platform for spike-based and rate-based models
hopefully triggers new research questions by facilitating collaboration
and translation of ideas between scientists working in the two fields.
This work therefore contributes to the unification of both modeling
routes in multi-scale approaches combining large-scale spiking networks
with functionally inspired rate-based elements to decipher the dynamics
of the brain.

\section{Appendix}

\subsection{Numerical evaluation of the Siegert formula}

We here describe how to numerically calculate \prettyref{eq:siegert_2D},
frequently called Siegert formula in the literature \citep[for a recent textbook see][]{Gerstner14},
which is not straight forward due to numerical instabilities in the
integral. First we introduce the abbreviations $y_{\theta}=(\theta-\mu)/\sigma+\gamma\sqrt{\taus/\taum}$
and $y_{r}=(\Vr-\mu)/\sigma+\gamma\sqrt{\taus/\taum}$ and rewrite
the integral as

\begin{align*}
 & \sqrt{\pi}\int_{y_{r}}^{y_{\theta}}e^{u^{2}}(1+\mathsf{erf}(u))du\\
= & 2\int_{y_{r}}^{y_{\theta}}\;e^{u^{2}}\int_{-\infty}^{u}e^{-v^{2}}\;dv\;du\\
= & 2\int_{y_{r}}^{y_{\theta}}\;\int_{-\infty}^{u}e^{(u+v)(u-v)}\;dv\;du\,.
\end{align*}
Here, the numerical difficulty arises due to a multiplication of a
divergent ($e^{u^{2}})$ and a convergent term ($1+\erf(u)$) in the
integrand. We therefore use the variable transform $w=v-u$ and obtain

\begin{align*}
= & 2\int_{y_{r}}^{y_{\theta}}\;\int_{-\infty}^{u}e^{(u+v)(u-v)}\;dv\;du\\
\stackrel{}{=} & 2\int_{y_{r}}^{y_{\theta}}\;\int_{-\infty}^{0}e^{(2u+w)(-w)}\;dw\;du\\
= & \int_{0}^{\infty}e^{-w^{2}}\frac{e^{2y_{\theta}w}-e^{2y_{r}w}}{w}\;dw\,,
\end{align*}
where we performed the integral over $u$ in the last line. 

For $y_{r},y_{\Theta}<0$ the integrand can be integrated straightforwardly
as
\begin{equation}
\int_{0}^{\infty}\frac{e^{2y_{\Theta}w-w^{2}}-e^{2y_{r}w-w^{2}}}{w}\;dw\ ,\label{eq:siegert2}
\end{equation}
where the two terms in the integrand converge separately and where,
in approximation, the upper integration bound is chosen, such that
the integrand has dropped to a sufficiently small value (here chosen
to be $10^{-12}$). Here, for $w=0$, the integrand has to be replaced
by $\lim_{w\rightarrow0}\frac{e^{2y_{\Theta}w}-e^{2y_{r}w}}{w}=2\left(y_{\Theta}-y_{r}\right)$.

For $y_{\theta}>0$ and $y_{r}<0$ only the combination of the two
terms in \prettyref{eq:siegert2} converges. So we rewrite
\begin{align}
 & \int_{0}^{\infty}e^{-w^{2}}\frac{e^{2y_{\Theta}w}-e^{2y_{r}w}}{u}\;dw\nonumber \\
= & \int_{0}^{\infty}e^{2y_{\theta}w-w^{2}}\frac{1-e^{2(y_{r}-y_{\Theta})w}}{w}\;dw\nonumber \\
= & e^{y_{\Theta}^{2}}\int_{0}^{\infty}e^{-(w-y_{\Theta})^{2}}\frac{1-e^{2(y_{r}-y_{\Theta})w}}{w}\;dw\,.\label{eq:siegert1}
\end{align}

The integrand has a peak near $y_{\Theta}.$ Therefore, in approximation,
the lower and the upper boundary can be chosen to be left and right
of $y_{\Theta}$, respectively, such that the integrand has fallen
to a sufficiently low value (here chosen to be $10^{-12}$). For $w=0$
we replace the integrand by its limit, which is $\lim_{w\rightarrow0}e^{-(w-y_{\Theta})^{2}}\frac{1-e^{2(y_{r}-y_{\Theta})w}}{w}=e^{-y_{\Theta}^{2}}2\left(y_{\Theta}-y_{r}\right)$.

We actually switch from \prettyref{eq:siegert2} to \prettyref{eq:siegert1}
when $y_{\theta}>0.05|\tilde{V}_{\theta}|/\sigma_{\alpha}$ with $\tilde{V}_{\theta}=V_{\theta}+\gamma\sqrt{\taus/\taum}$.
This provides a numerically stable solution in terms of a continuous
transition between the two expressions. Our reference implementation
numerically evaluates the integrals using the adaptive GSL implementation
\texttt{gsl\_integration\_qags} \citep{GSL06} of the Gauss-Kronrod
21-point integration rule. 

\section*{Conflict of Interest Statement\pdfbookmark[1]{Conflict of Interest Statement}{ConflictsPage}}

The authors declare that the research was conducted in the absence
of any commercial or financial relationships that could be construed
as a potential conflict of interest.

\section*{Acknowledgments\pdfbookmark[1]{Acknowledgments}{AcknowledgmentsPage}}

The authors gratefully acknowledge the computing time on the supercomputer
JUQUEEN \citep{stephan2015juqueen} at Forschungszentrum Jülich granted
by firstly the JARA-HPC Vergabegremium (provided on the JARA-HPC partition,
jinb33) and secondly by the Gauss Centre for Supercomputing (GCS)
(provided by the John von Neumann Institute for Computing (NIC) on
the GCS share, hwu12). This project has received funding from the
European Union's Horizon 2020 research and innovation programme under
grant agreement No 720270 (HBP SGA1). Partly supported by Helmholtz
Portfolio Supercomputing and Modeling for the Human Brain (SMHB),
the Initiative and Networking Fund of the Helmholtz Association, the
Helmholtz young investigator group VH-NG-1028, and the Next-Generation
Supercomputer Project of MEXT. All network simulations carried out
with NEST (http://www.nest-simulator.org).

\bibliographystyle{neuralcomput_natbib}
\phantomsection\addcontentsline{toc}{section}{\refname}\bibliography{brain,computer,math}

\end{document}